\documentstyle[11pt,aaspp4]{article}
\newcommand{\iras}{{\sl IRAS\/}}
\newcommand{\etal}{{\sl et al.\/}}
\newcommand{\mpc}{{$h^{-1}$ Mpc}}
\def\ltsima{$\; \buildrel < \over \sim \;$}
\def\lsim{\lower.5ex\hbox{\ltsima}}
\def\gtsima{$\; \buildrel > \over \sim \;$}
\def\gsim{\lower.5ex\hbox{\gtsima}}

\accepted{19 August 1997}
\lefthead{Kim \& Strauss}
\righthead{High-Order Moments of the Galaxy Distribution}

\begin{document}

\title{Measuring High-Order Moments of the Galaxy Distribution from
Counts in Cells -- The Edgeworth Approximation }

\author{Rita Seungjung Kim and Michael A. Strauss\altaffilmark{1}}
\authoremail{(rita,strauss)@astro.princeton.edu}
\affil{Department of Astrophysical Sciences, Peyton Hall, Princeton
University, Princeton, NJ 08544}
\altaffiltext{1}{Alfred P. Sloan Foundation Fellow}

\begin{abstract}

To probe the weakly non-linear regime, past the point where simple
linear theory is sufficient to describe the statistics of the density
distribution, we measure the skewness ($S_3$) and kurtosis ($S_4$) of
the Count Probability Distribution Function (CPDF) of the \iras\ 1.2
Jy sample obtained from counts in cells.  These quantities are free
parameters in a maximum likelihood fit of an Edgeworth expansion
convolved with a Poissonian to the observed CPDF.  This method,
applicable on scales $\gsim5$\mpc, is appreciably less sensitive to
the tail of the distribution than are measurements of $S_3$ and $S_4$
from moments of the CPDF.  We measure $S_3$ and $S_4$ to $l\sim 50
h^{-1}\mbox{Mpc}$; the data are consistent with scale invariance,
yielding averages of $\langle S_3 \rangle = 2.83\pm 0.09$, and
$\langle S_4 \rangle = 6.89 \pm 0.68$.  These values are higher than
those found by Bouchet \etal\ (1993) using the moments method on the
same data set, $\langle S_3 \rangle = 1.5\pm 0.5$ and $\langle
S_4\rangle = 4.4\pm 3.7$, due to lack of correction in the latter work
for finite-volume effects.  Unlike the moments method, our results are
quite robust to the fact that \iras\ galaxies are under-represented in
cluster cores.  We use $N$-body simulations to show that our method
yields unbiased results.
  
\end{abstract}

\section{Introduction}
\label{sec:intro}

Many approaches have been made to characterize the clustering of
galaxies, especially over the past decade as better and deeper
redshift surveys have become available (cf., Borgani 1995; Strauss \&
Willick 1995, for comprehensive reviews). The two-point correlation
function and its Fourier Transform, the power spectrum, the long
popular methods for describing the clustering of galaxies, are
complete statistical descriptions of the density field only if the
phases of the Fourier modes of the density field are random.  Indeed,
simple inflationary models predict these random phases; if this
condition holds, the one-point probability distribution function (PDF)
of the density field $\delta ({\bf r})$ is Gaussian.

As perturbations grow by gravitational instability, an initially
Gaussian distribution remains Gaussian as long as the fluctuations
remain in the linear regime.  However, once non-linear effects become
important, the distribution deviates from its initial Gaussian state,
and one needs higher order statistics to characterize the density
field.

For a zero-mean Gaussian distribution all reduced moments (cumulants)
of the PDF are zero except the variance ($\langle \delta^2 \rangle
\equiv \sigma^2$), hence non-zero skewness $\langle \delta^3 \rangle$,
kurtosis $\langle \delta^4 \rangle - 3\sigma^4 $, and higher order
cumulants are measures of the deviation of the distribution from a
Gaussian.  In this paper, we only consider the first two lowest order
effects, the skewness and the kurtosis.  These $N$th-order cumulants
are equal to the volume averaged correlation functions,
\begin{equation}
\bar{\xi}_N (v) = {1 \over v^N} \int_v d^3 {\bf r}_1 d^3 {\bf r}_2 \ldots
    d^3 {\bf r}_N \,\xi_N ({\bf r}_1 , {\bf r}_2 , \ldots ,{\bf r}_N ) ,
	\label{eq:avecorrel}
\end{equation}
where $\bar{\xi}_2 \equiv \sigma_2 \equiv\langle \delta^2 \rangle$ , 
$\bar{\xi}_3 \equiv \langle \delta^3 \rangle $, 
$\bar{\xi}_4 \equiv \langle \delta^4 \rangle - 3\sigma^4$
and so forth.  Here the volume $v$ over which the $\bar{\xi}_N$ are
averaged is defined by the smoothing scale of the density field
$\delta$.
 
The assumption of scale invariance by Balian \& Schaeffer (1988, 1989),
\begin{equation}
 \xi_N ( \lambda {\bf r}_1 , \ldots , \lambda {\bf r}_N ) = 
       \lambda^{\gamma (N-1) } \xi_N ({\bf r}_1 , \ldots , {\bf r}_N),
\end{equation}
yields the scaling relation, 
\begin{equation}
 \bar{\xi}_N (v) = S_N \bar{\xi}_2^{N-1} (v) , \label{eq:scaling}
\end{equation}
where the $S_N$ are independent of scale.  The scale invariance of
$S_N$ and the scaling relation (\ref{eq:scaling}) are in fact
predicted by perturbation theory in the mildly non-linear regime,
under the two assumptions of Gaussian initial conditions and growth of
conditions via gravitational instability (Fry 1984ab, Bernardeau
1992).  Thus one can test the scale-invariance model by measuring the
dependence of $S_N$ on smoothing scale, although it can be difficult
in practice to rule out non-Gaussian models (Fry \& Scherrer 1994;
Bouchet \etal\ 1995).

Calculation  of $S_N$ from gravitational instability  invokes
$(N-1)$th order perturbation theory.  Bernardeau (1994b) 
presents a method for calculating $S_N$ for top-hat filters; the
results are:
\begin{equation}
	S_3 = {34 \over 7} - (n +3) ,  \label{eq:s3perturb}
\end{equation}
and
\begin{equation}
	S_4 = {60712 \over 1323} - {62\over 3} (n+3) +
		{7\over3}(n+3)^2, \label{eq:s4perturb}
\end{equation}
where $n$ is the spectral slope of the power spectrum; a pure
power-law spectrum is assumed.  The expressions above are for
$\Omega_0 = 1$; while the $S_N$ are sensitive to the slope of the
power spectrum, the dependence on $\Omega_0$ is quite weak (Bouchet
\etal\ 1992; 1995).  These, and the corresponding results for a
Gaussian filter have been confirmed with $N$-body simulations
(Juszkiewicz, Bouchet, \& Colombi 1993; Bernardeau 1994b; {\L}okas
\etal\ 1995; Juszkiewicz \etal\ 1995).  On the observational side,
calculations of $S_3$ and $S_4$ have been done for the CfA (Huchra
\etal\ 1983) and SSRS (da Costa \etal\ 1991) catalogs by Gazta\~naga
(1992) and Fry \& Gazta\~naga (1994), and on the \iras\ 1.2 Jy sample
(Fisher \etal\ 1995) by Bouchet \etal\ (1993, hereafter B93).
Calculations of higher order angular moments have been done for the
Lick galaxy counts (Szapudi, Szalay, \& Boschan 1992), \iras\ galaxies
(Meiksin, Szapudi, \& Szalay 1992), the APM galaxy survey (Gazta\~naga
1994, 1995; Szapudi \etal\ 1995), and the EDSGC (Szapudi, Meiksin, \&
Nichol 1996).  For optically selected galaxies, Gazta\~naga (1992)
found that $S_3 = 1.94 \pm 0.07$ up to a smoothing scale of $\sim$ 22
\mpc, which is slightly higher than the value found by B93 for the
\iras\ sample: $S_3 = 1.5 \pm 0.5$.

The standard technique to measure $S_3$ and $S_4$ from observational
data involves calculation of the moments of the Count Probability
Distribution Function (CPDF)\footnote{The PDF refers to the
distribution function of the underlying continuous density field
$\delta$, while the CPDF is the distribution function of the
discretely sampled galaxy distribution.} (\S~\ref{sec:moments}), which
in turn is determined via counts in cells.  The high-order moments of
the CPDF are of course weighted heavily by its high density tail.
Regions of such high density are rare, so these moments are highly
sensitive to the presence or absence of a few clusters (Colombi,
Bouchet, \& Schaeffer 1994, 1995; Szapudi \& Szalay 1996).  Also, in a
finite volume, there is always a densest region, and thus the CPDF
goes to zero for higher densities.  Not taking this finite-volume
effect into account will cause the clustering amplitudes to be
systematically underestimated.  The CPDF asymptotes to an exponential
at high densities, especially in the strongly non-linear regime
(Balian \& Schaeffer 1989); one can thus extrapolate the observed CPDF
to arbitrarily high densities.  Thus one can obtain unbiased estimates
of the high-order moments, {\em if the volume is large and dense
enough to reach this asymptotic regime, to allow the exponential to be
fit\/} (Fry \& Gazta\~naga 1994; Colombi \etal\ 1994, 1995).  This
exponential asymptotic behaviour is not expected for the weak regime,
making it difficult to correct for finite-volume effects (cf., the
discussion in B93).  Finally, the tail of the distribution is also
affected by {\it finite-sampling\/} effects; it can be underestimated
if the CPDF is determined from too few spheres (Szapudi \& Colombi
1996).

However, as we shall see, skewness and kurtosis affect the entire
CPDF, not just the tails.  In particular, skewness causes the mode of
the distribution (the region where the measured CPDF is most robust)
to shift from the mean.  This motivates us to develop a new method to
measure the $S_N$ from the CPDF, less sensitive to finite volume
effects, by fitting the entire CPDF to a functional form.

There are several approaches to calculating the evolution of the PDF
of $\delta$ from Gaussian initial conditions, using the Zel'dovich
(1970) approximation (Kofman \etal\ 1994) or Eulerian perturbation
theory (Bernardeau 1992; Bernardeau \& Kofman 1995; Colombi \etal\
1997).  The so-called Edgeworth expansion, which provides a convenient
parametric form to account for small deviations from Gaussianity,
gives an excellent fit to the PDF of $\delta$ in $N$-body models for
small $\sigma$ (Juszkiewicz \etal\ 1995).  In this paper, we take the
Edgeworth expansion convolved with a Poisson distribution as our
model, and perform a maximum likelihood fit with respect to the free
parameters $S_N$ to the observed CPDF of the \iras\ 1.2 Jy survey from
B93.  We expect this method to be more robust than direct calculation
of the moments, since it depends more on the overall shape of the
distribution function than on the high-density tail region.  Although
there are several applications of the Edgeworth expansion to measure
non-Gaussian statistics of the density field in the literature
(Scherrer \& Bertschinger 1991; Amendola 1994; Juszkiewicz \etal\
1995), it has not yet been applied to observational data.  We check
the validity of our technique by applying it to \iras\ mock catalogs
taken from $N$-body simulations, and compare the results with the
predicted value of $S_3$ and $S_4$ from perturbation theory, and with
values measured from the moments of the PDF.

In \S 2 we give a brief account of the moments method, and describe
our model based on the Edgeworth expansion.  We test this method with
$N$-body simulations in \S 3.  In \S 4, we apply our method to the
\iras\ 1.2 Jy CPDF and compare our results with those of the moments
method and perturbation theory.  We summarize our results in \S 5.

\section{Method and Analysis}

\subsection{Count Probability Distribution Function and its Moments}
\label{sec:moments}

The CPDF $P_N (l)$ is defined as the fraction of randomly positioned
spheres of radius $l$ containing exactly $N$ galaxies for a given
volume-limited galaxy sample.  Here we use the CPDF of \iras\ galaxies
from 10 volume-limited subsamples as calculated by B93 (see their
Table 1).  We place $10^6$ random points in each subsample and count
the number of galaxies in concentric spheres of different radii $l$,
from each point, considering only those spheres that are completely
included within the subsamples (see B93 for details).

B93 calculate the normalized cumulants $S_N$ by the {\it moments
method}.  The moments of the distribution $P_N(l)$ are given by
\begin{equation}
	\mu_M (l)  = 
	\left\langle \left( {N - \bar{N}}\over \bar{N} \right)^M \right\rangle
	= \sum_{N=0}^{\infty} \left( {N - \bar{N}} \over \bar{N} \right)^M
		P_N (l)  ,
\end{equation}
where $ \bar{N} \equiv \langle N \rangle = \sum N P_N (l)$ is the mean
number of galaxies in a sphere of size $l$.  The first few
volume-averaged correlation functions (reduced moments), corrected for
shot noise are given by (Peebles 1980)
\begin{eqnarray}
	\bar{\xi}_2 (l)  &=&  \mu_2 - {1 \over \bar{N}} , \\
	\bar{\xi}_3 (l)  &=&  \mu_3 - 3 {\mu_2 \over \bar{N}} + {2\over \bar{N}^2} ,\\
	\bar{\xi}_4 (l)  &=&  \mu_4 - 6 {\mu_3 \over \bar{N}} - 3\mu_2^2 
	 	   + 11 {\mu_2 \over 	\bar{N}^2} - {6\over \bar{N}^3} . 
\end{eqnarray}
The skewness and kurtosis, $S_3 \ \mbox{and} \ S_4$, then follow from
equation~(\ref{eq:scaling}).  These calculations were done by B93 for
the \iras\ redshift survey, and it was found that the scaling relation
(equation~\ref{eq:scaling}) indeed holds very well (see Figure~8 of
B93).  A fit of the data to $\log \bar{\xi}_N = C_N + D_N \log
\bar{\xi}_2$, gives $ D_3 = 1.96 \pm 0.06$ and $D_4 = 3.03 \pm 0.18$,
where scale invariance predicts $D_N = N-1$
(equation~\ref{eq:scaling}).  All calculations are done in redshift
space, but the $S_N$ are quite insensitive to redshift space
distortions, at least on mildly non-linear scales (Bouchet \etal\
1992; Lahav \etal\ 1993; Fry \& Gazta\~naga 1994; Hivon \etal\ 1995).
As mentioned in the previous section, no correction for finite-volume
or finite-sampling effects have been carried out for these data.

\subsection{The Edgeworth Expansion} 
\label{sec:edgeworth}

The primordial density fluctuations are assumed to be Gaussian
distributed, and as these fluctuations grow by gravitational
instability, the PDF of $\delta$ deviates away from its initial
Gaussian form, generating non-zero higher order moments.  To the
extent that the deviations from a Gaussian are small, it makes sense
to write the PDF as an expansion around a Gaussian. The Edgeworth
expansion is a rigorous way to do this, as described by Juszkiewicz
\etal\ (1995).

We expand the PDF, here denoted by $p(\nu )$, where $\delta \equiv
(\rho - \bar{\rho}) / \bar{\rho}$, $\nu \equiv \delta / \sigma$, in
terms of a Gaussian
\begin{equation}
	\phi (\nu) = {1\over \sqrt{2\pi}} \exp (- \nu^2 / 2 )
\end{equation}
and its derivatives. This is called the Gram-Charlier series (Cram\'er
1946),
\begin{equation}
	p(\nu) = \phi (\nu) \left[c_0  - {c_1 \over 1!} H_1(\nu) +
		 {c_2 \over 2!} H_2(\nu) + \cdots \right] ,
\label{eq:gram-charlier} 
\end{equation}
where the $H_l$ are the Hermite polynomials, as given in Table 1.
By the orthogonality of the $H_l$ one obtains,
\begin{equation}
	c_l = (-1)^l \int_{-\infty}^{\infty} H_l (\nu) p(\nu) d\nu .
\end{equation}
Therefore the first few coefficients of equation~(\ref{eq:gram-charlier})
are given by:
\begin{equation}
	c_0 = 1, \quad c_1=c_2=0, \quad c_l = (-1)^l S_l \sigma^{l-2} 
	\quad (3 \leq l\leq 5),\quad c_6 = S_6 \sigma^4 + {S_3}^2 \sigma^2
\end{equation}
where the $S_l$ are the normalized cumulants defined in
equation~(\ref{eq:scaling}).  A reordering of the terms of the
Gram-Charlier series gives a proper asymptotic expansion in $\sigma$,
the Edgeworth series:
\begin{equation}
	p(\nu) = \phi(\nu)\left\{1 + {1\over 3!} S_3 H_3(\nu) \sigma
		 + \left[{1\over 4!} S_4 H_4(\nu)
		 + {10 \over 6!} S_3^2 H_6(\nu)\right] \sigma^2 +
{\cal O}(\sigma^3) \right\}. 
	\label{eq:fullew}
\end{equation}
\placetable{table:1}

\subsection{A Model for the CPDF}
\label{sec:analysis}

Equation~(\ref{eq:fullew}) is a model for the underlying density
distribution from which the galaxies are sampled, but it does not yet
account for the discreteness of the galaxies.  Due to the finite
number of galaxies in each sample, the observed CPDF is subject to
Poisson noise, and we take this effect into account by convolving the
Edgeworth expansion with a Poisson distribution (cf., Coles \& Jones
1991).

We define the density contrast as $\delta \equiv {N - \langle N
\rangle \over \langle N \rangle} $.  Let us rewrite the Edgeworth
expansion to third order (second order in $\sigma$) as a function of
$\delta$,
\begin{equation}
	E(\delta) = {1\over \sqrt{2\pi} \sigma} e^{-\delta^2 / 2\sigma^2 }
		\left\{ 1 + {1\over 6} S_3 \sigma H_3(\delta / \sigma ) 
		       + \left[ {S_4 \over 24} H_4 (\delta/\sigma ) 
				+ {S_3^2 \over 72} H_6 (\delta /\sigma )
			 \right] \sigma^2 
		\right\}.
	       				\label{eq:edelta3}
\end{equation} 
The expectation value of $P(N)$ at a given value of $l$ is given by
the convolution of equation~(\ref{eq:edelta3}) with a Poisson
distribution,
\begin{equation}
	 \langle P(N) \rangle = \int d\delta\, E(\delta) 
		      F( N\! \mid\! N_{\delta}\!\equiv\! \langle N \rangle
					(\delta +1) )  \label{eq:convolve}
\end{equation}
where the Poisson distribution is 
\begin{equation}
 	 F(N \mid N_{\delta}) 
			= {{N_{\delta}} ^N \over N!}  e^{-N_{\delta}},
\end{equation}
the conditional probability of finding $N$ points in a sphere when the
true overdensity in that sphere is $\delta = (N_\delta - \langle N
\rangle) / \langle N \rangle $.

\begin{figure}[p]
\plotfiddle{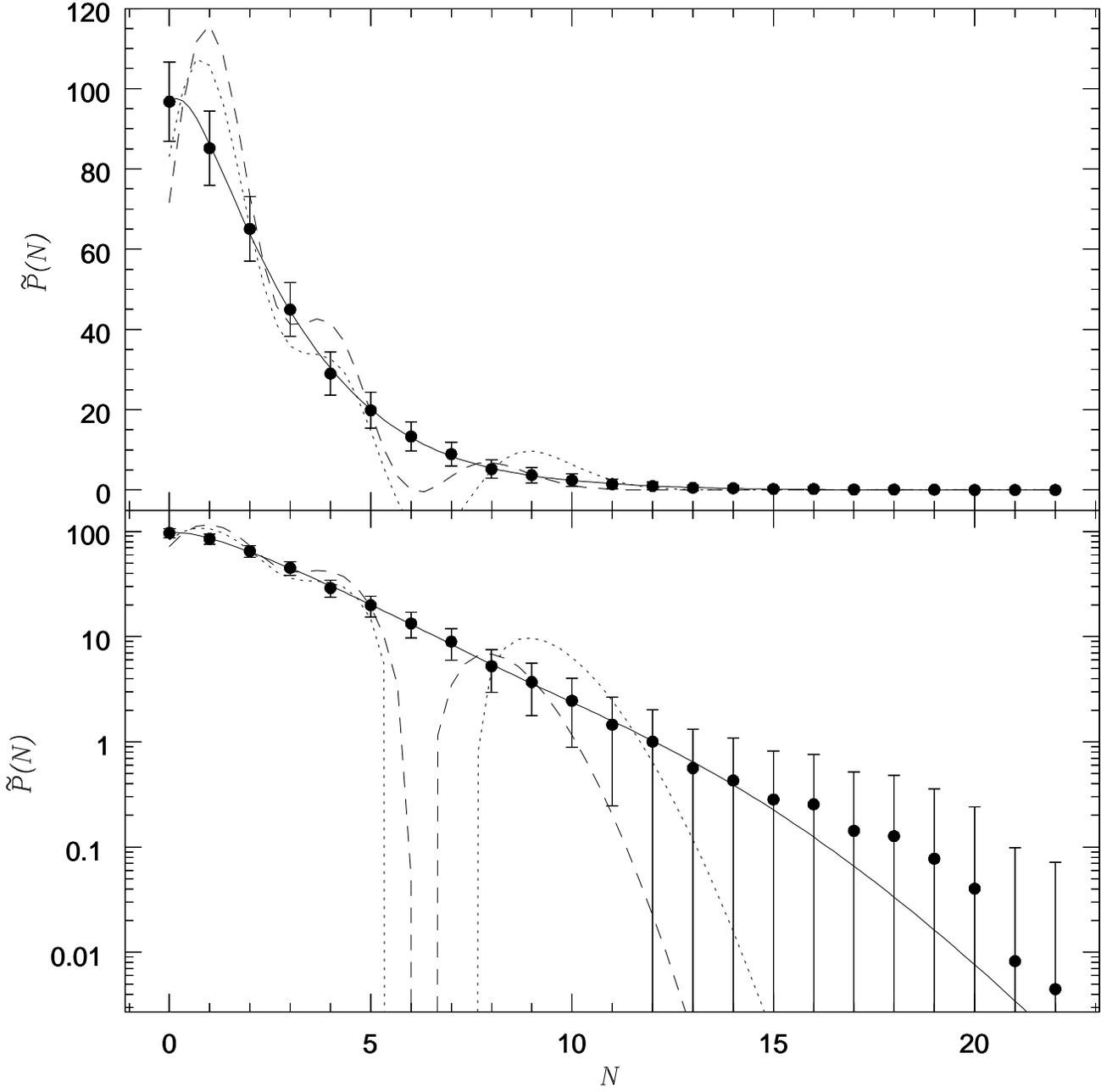}{5.8in}{0}{90}{90}{-290}{-140}
\caption{ 
Comparison of the fit of the Edgeworth expansion convolved
with a Poissonian (solid line) to the \iras\ CPDF of sample size $R=
59\, h^{-1}$Mpc and smoothing length $l = 7.92\, h^{-1}$Mpc (dots),
and the corresponding underlying density field (the pure Edgeworth
expansion without shot noise component; dashed line).  The top panel
is a linear plot, while the bottom has a logarithmic y-axis to better
show the tail.  The best-fit values of the parameters are $\sigma =
0.77,\, S_3 = 2.56,\, S_4 = 7.38$.  The dotted line is the result of a
pure Edgeworth expansion fitted to the CPDF, without convolution with
a Poissonian; the resulting parameters are $\sigma = 0.91,\, S_3 =
2.92,\, S_4 = 12.95$. 
\label{fig:underlying}
	}
\end{figure}

The Edgeworth expansion is not positive definite; moreover, for values
of $\sigma$ approaching unity, it shows unphysical oscillations
(Juszkiewicz \etal\ 1995; Ueda \& Yokoyama 1996).  However, when we
convolve it with a Poisson distribution, the Edgeworth expansion
becomes much better behaved. Figure~\ref{fig:underlying} illustrates
this with the observed CPDF for a volume-limited sample of \iras\
galaxies (points).  The CPDF is normalized to the total number of
spheres which fill the volume, $M$ (equation~\ref{eq:m-def}), and the
error bars are given by Poisson statistics, i.e., the square root of
the value of the CPDF.  The solid curve is the best fit of the model
in equation~(\ref{eq:convolve}) (using the method described below),
while the dashed curve shows the underlying Edgeworth expansion
(equation~\ref{eq:edelta3}) with the same values of $\sigma$, $S_3$,
and $S_4$.  The dashed line goes negative and oscillates around the
CPDF, while the solid line traces it nearly perfectly. This example
has $\sigma = 0.77$.  The dotted line is the result of performing a
fit of the Edgeworth expansion with no Poisson noise term included.
Not surprisingly, the $\sigma$ is overestimated, and the model gives a
very poor fit.  Thus for the sparse \iras\ data we use in this paper,
the Poisson noise term is absolutely essential in our model for the
CPDF.  Note that the Edgeworth expansion is no longer valid for
$\sigma > 1$, and thus the example in Figure~\ref{fig:underlying}
represents the smallest scales on which we will apply it.


\begin{figure}[p]
\plotfiddle{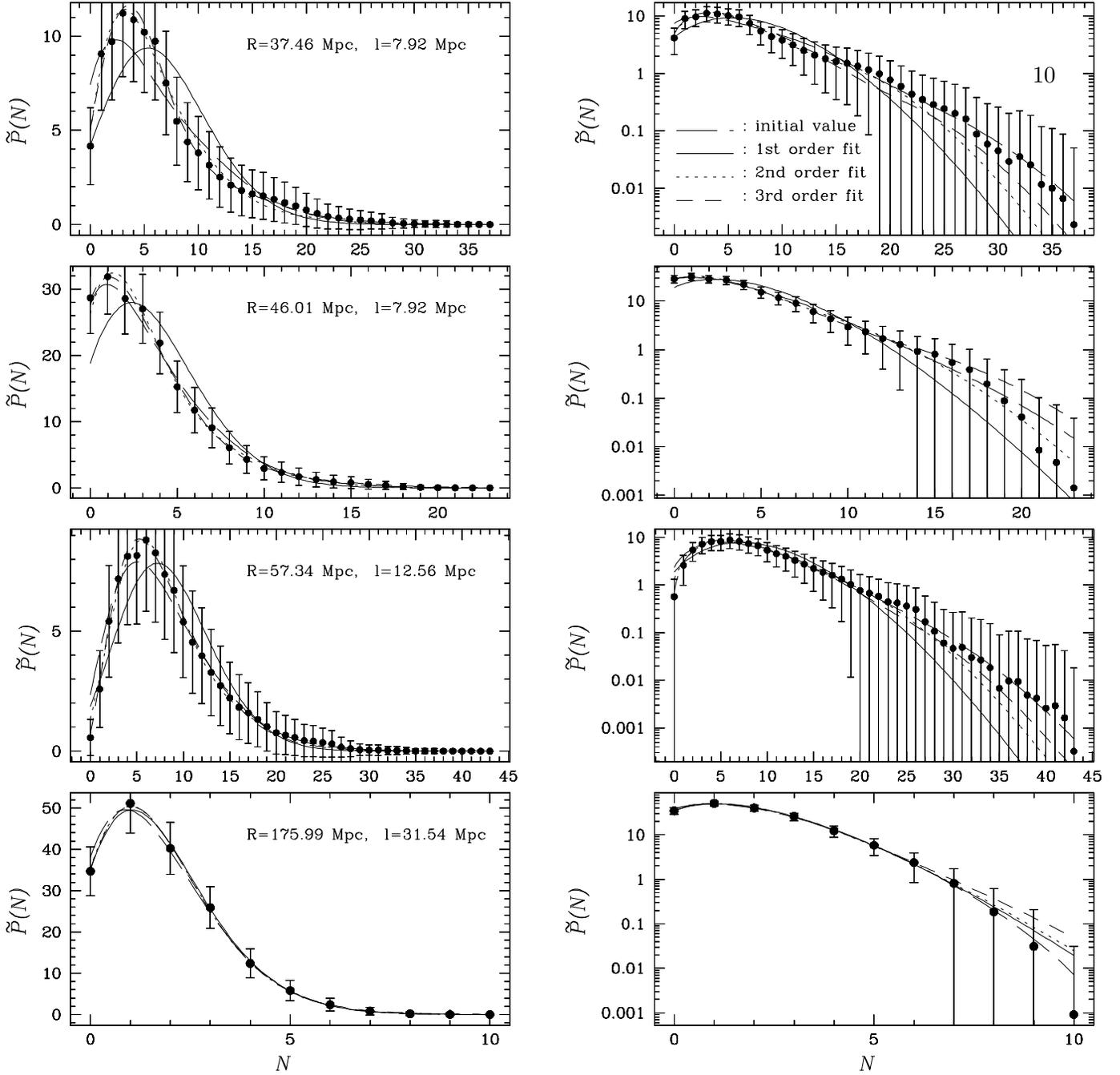}{6.1in}{0}{100}{100}{-310}{-160}
\caption{
Various fits of the Edgeworth expansion convolved with a Poissonian to
the observed CPDF (denoted by dots), with error bars given by Poisson
statistics. $R$ is the size of the \iras\ sample, and $l$ the
smoothing scale. The long dashed line represents the model using the
$\sigma$, $S_3$, and $S_4$ of the moments method, the solid line our
first order fit, the dotted line the second and short dashed the third
order fit. The volume of the subsample increases as one goes
downwards, and the solid line (Gaussian) becomes a better
approximation.  The right panels use a logarithmic y-axis to show the
tails.  The fits trace the data points reasonably well even when
$\tilde{P}(N) < 1$, which were excluded in the likelihood analysis. 
\label{fig:4datafits}
	}
\end{figure}

The observed CPDF, $P(N)$, and the model, $\langle P(N) \rangle$, are
defined to be the {\it probability\/} that a sphere of size $l$
contains $N$ points.  When we assume that a total of $M$ spheres are
placed randomly within the volume, then the {\it number\/} of spheres
that contain exactly $N$ points is $\tilde{P}(N) \equiv M P(N)$; we
similarly define $\langle \tilde{P}(N) \rangle \equiv M \langle P(N)
\rangle $.  We would like $M$ to represent the number of {\em
statistically independent\/} spheres, in order to allow us to define
error bars on the observed $P(N)$, but it is not clear {\it a
priori\/} how to measure this.  The number of spheres thrown within
the volume, $10^6$, is clearly an overestimate for $M$, due to severe
overlap between spheres\footnote{$10^6$ may in fact be too {\it
small\/} a number of spheres to avoid finite-sampling effects in the
tails of the distribution; cf., Szapudi \etal\ (1996) and Szapudi \&
Colombi (1996).  However, we argue below that the Edgeworth method is
insensitive to such biases in the tail.}.  One possibility, which is
used to define errorbars in Figures~\ref{fig:underlying} and
\ref{fig:4datafits}, is to take
\begin{equation} 
	M = {\omega\over {4\,\pi}}\left( {R \over l} \right)^3,
\label{eq:m-def}
\end{equation}
the ratio of the volumes of the sample and the sphere\footnote{Note
that $M$, and accordingly $\tilde P(N)$, are not integers.}; here $R$
is the radius of the subsample, and $\omega$ is the solid angle it
subtends.  However, Gazta\~naga \& Yokoyama (1993) show (as we confirm
in \S~\ref{sec:errors}) that this is an underestimate, and suggest
multiplying equation~(\ref{eq:m-def}) by $\sigma^{-3}$ (which is much
larger than unity on large scales).  This issue is further discussed
in Szapudi \& Colombi (1996) and Szapudi \etal\ (1996).  We do not
have a rigorous solution to this problem. Our approach for the present
paper is to use the value of $M$ given by equation~(\ref{eq:m-def}),
demonstrate directly that it is an underestimate, and, in the
following subsection, suggest an empirical rescaling to allow us to
define error bars on measured quantities.

As explained above, we throw $10^6$ spheres in each volume to measure
$P(N)$.  This is multiplied by $M$, which is several orders of
magnitude lower than $10^6$, to obtain $\tilde{P}(N)$.  We assume that
this {\it number} of spheres $\tilde{P}(N)$ is Poissonian distributed
around the true value $\langle \tilde P(N) \rangle$, so the likelihood
of each observed data point $\tilde{P}(N)$ is given by the Poisson
distribution.  We assume that the values of $\tilde P(N)$ are
statistically independent (We will see in the following subsection
that these assumptions appear to be violated, but that we can make a
heuristic fix to the likelihood).  We therefore can express the
likelihood function of the observed CPDF as the product of these
quantities over $N$,
\begin{equation}
	{{L}} = \prod_N {{\langle \tilde{P} (N) \rangle^{\tilde{P}(N)}}
 		\over {\tilde{P}(N) !}}\, e^{-\langle \tilde{P} (N)\rangle}.
					\label{eq:likelihood}
\end{equation}
In practice, the product extends only over those values of $N$ for
which $\tilde{P}(N) > 1$, as the Poisson model breaks down beyond
that.  Nevertheless, we will see that in many cases the best fitting
curve continues to trace the CPDF reasonably well even when
$\tilde{P}(N) < 1$ (cf. the right-hand panels of
Figure~\ref{fig:4datafits}).  We have assumed a Poisson distribution
in $\tilde{P}(N)$; if we had assumed a Gaussian distribution, our
likelihood function could be expressed in terms of $\chi^2$. Here, the
quantity corresponding to $\chi^2$ is then:
\begin{equation}
	{\cal L} = -2 \ln L = \sum_N
	2\left[ \langle \tilde{P}(N)\rangle
		-\tilde{P}(N)\ln\langle\tilde{P}(N)\rangle
		+\ln (\tilde{P}(N)!)	
	\right],		\label{eq:chisquared}
\end{equation}	 		
and we minimize this quantity instead of maximizing $L$
(equation~\ref{eq:likelihood}).  We use Powell's method (Press \etal\
1992) to minimize ${\cal L}$ with respect to $\sigma,\ S_3\
\mbox{and}\ S_4$; indeed, we do fits to first order (i.e., Gaussian),
second order (including the term proportional to $S_3 \sigma$) and
third order in the Edgeworth expansion (equation~\ref{eq:edelta3}),
although we focus mainly on the third-order fit in this paper.  To
avoid having the code settle into a local minimum, we start the search
with reasonable values of the parameters, i.e., those we calculate by
the moments method.  In most cases this converges to a fair estimate
of the parameters, but in some cases, especially on large scales,
where the higher order correlations are intrinsically weak, or when
the volume of the subsample is very large and hence the density of the
galaxies are small, calculating the moments yields negative quantities
of $S_3$ and $S_4$.  When this happens, we set the initial values of
$S_3$ and $S_4$ to zero instead, and minimize ${\cal L}$ again.


Maximum likelihood fits of first, second and third order to the \iras\
CPDF's for representative values of $R$ and $l$ are given in Table~2;
those with asterisks are plotted in Figure~\ref{fig:4datafits}.  The
left panels of the figure give $\tilde{P}(N)$ on a linear scale to
show the fit around the peak clearly. The right panels are log plots
to show the tail region in detail.  The error bars are again given by
Poisson statistics as in Figure~\ref{fig:underlying}.  The solid line
is the first order (Gaussian) fit, and the dotted line and the short
dashed line represent second and third order fits respectively. The
second and third order fits are a remarkable improvement over the
initial set of parameters obtained by the moments method, represented
by the long dashed lines.  The third order fits, with the term
proportional to $\sigma^2$ included, are an improvement over the
second order fits.  Note in particular that the third-order fits
follow the observed CPDF even into the region where $\tilde P(N) < 1$,
which is not even included in the fitting procedure.  On the other
hand, quite interestingly, the third order model with the parameters
given by the moments method (long dashes) usually does the best job of
fitting the high-density tail of the CPDF, although this fit is poor
at intermediate to low $\delta$.  This again reflects the sensitivity
of the moments method to the tails of the distribution.  As the volume
of the subsample increases, going downward in the figure, the Gaussian
fit becomes a better approximation to the observed CPDF.  Indeed, for
the bottom-most panel with a subsample of volume 176 \mpc and a
smoothing length of 31.54 \mpc, the first, second, and third order
fits are barely distinguishable. This is a result of two
effects. First, the larger the subsample is, the sparser the sample
becomes, and therefore the Poisson noise dominates the CPDF, swamping
the higher order correlations. Second, the larger the smoothing scale,
the weaker the higher order correlations become, because clustering is
weaker at larger scales. Therefore in either of these cases, the
accuracies of the higher order correlation terms, $S_3$ and especially
$S_4$, drop appreciably, and the error bars we derive for these
quantities in the next section are therefore quite large.

\subsection{Error Estimates and the Value of $M$}
\label{sec:errors}

We determine errors on the parameters from the inverse of the Hessian
matrix of the likelihood function. Close to the minimum, we expect the
likelihood function to be well approximated by a quadratic form,
\begin{equation}
	{\cal L}({\bf x}) \approx {\cal L}({\bf x_0}) + {1\over 2} ({\bf x}-{\bf x_0}) \cdot {\bf D} 
		\cdot ({\bf x}-{\bf x_0}),
\end{equation}
where ${\bf x}$ is the vector of parameters, and ${\bf x_0}$ is the
value of this vector at the minimum value of $\cal L$. Here ${\bf D}$,
the Hessian matrix, is the second derivative matrix of ${\cal L}$.
Since the form of the likelihood function is given, the Hessian matrix
is known to us. The covariance matrix is then obtained by
\begin{equation}
	[C] \equiv 2\,{\bf D}^{-1} 
\end{equation}
and $\sqrt{C_{ii}}$ are the $1\sigma$ confidence intervals of the
parameters  $x_i$.

The values of the errors derived in this way are of course critically
dependent on the value of $M$ assumed.  The third-order model fits the
points in Figures~\ref{fig:underlying} and \ref{fig:4datafits} much
too well, given the Poisson errors shown, suggesting that our value of
$M$ is underestimated.  We can quantify this with a $\chi^2$
statistic:
\begin{equation} 
\chi^2 = \sum_N {{(\tilde P(N) - \langle \tilde P(N) \rangle)^2}\over
\tilde P(N)},
\label{eq:chi2} 
\end{equation}
where Poisson error bars are assumed, and the sum is over only the
${\cal N}$ values of $N$ for which $\tilde P(N) > 4$ (such that the
correspondence between the Poisson distribution, and the Gaussian
distribution assumed in equation~(\ref{eq:chi2}), is valid).  This
quantity is less than the number of degrees of freedom ${\cal N} - 4$
by factors of as much as ten.  As we discussed in the previous
subsection, it is not clear how to set $M$ {\it a priori\/}, and
consequently the errors on $P(N)$ and parameters derived from it (cf.,
the exhaustive discussion in Szapudi \& Colombi 1996 for the strongly
non-linear regime; there does not yet exist a complete theory for the
errors in the weakly nonlinear regime).  We find that the Gazta\~naga
\& Yokoyama (1993) suggestion of multiplying $M$ by $\sigma^{-3}$ does
not bring $\chi^2/({\cal N} - 4)$ to unity, although it does go in the
right direction.

Our approach is an empirical one: we {\it scale\/} our errors of
$\tilde P(N)$ to force $\chi^2 = {\cal N} - 4$; equivalently, we
multiply the errors in $\sigma$, $S_3$, and $S_4$ derived above by
$(\chi^2/({\cal N} - 4))^{1/2}$.  This is done {\it a posteriori}, and
so does not affect the best-fit values of the parameters, but it does
of course have a strong effect on the derived errors.  This is not a
statistically rigorous procedure, but we will justify these errors
empirically in the next section.  The $1\, \sigma$ errors determined
in this way are included in Table~2.

Table~2 includes the values of $\cal L$ for each fit; these have been
scaled by the factor $\chi^2/({\cal N} - 4)$.  This way one can
quantify the goodness of fit of the curves seen in
Figure~\ref{fig:4datafits} by comparing the relative likelihoods. The
lower the value of ${\cal L}$, the better the fit becomes for given
values of $R$ and $l$; a difference of unity is significant at the
1$\,\sigma$ level.  These values of ${\cal L}$ are not comparable
across subsamples or smoothing scales, since in each case the input
CPDF is different.

\section{Tests with $N$-Body Simulations}
\label{sec:nbodytest}

To check the validity of our method, we generate \iras\ mock catalogs
from the $N$-body simulations of Protogeros \& Weinberg (1997), kindly
given to us by D. Weinberg.  These assume an initial power spectrum
$P(k) \propto k^{-1}$, and $\Omega_0 = 0.3$, and are evolved forward
to the point when $\sigma_8$, the rms fluctuation amplitude within an
8 \mpc\ sphere, is equal to 0.8.  The simulations use a staggered mesh
PM code by Park (1990), and are evolved within a cube of size 300
\mpc, containing $150^3$ particles and a $300^3$ mesh.  We assume that
the galaxy distribution is unbiased relative to the dark matter.  We
choose a random point within the simulation to represent the Local
Group, and produce 10 concentric volume-limited mock catalogs with
exactly the same volumes and number densities as in the real universe
(cf., Table 1 of B93).  We compute the CPDF for these samples with
tophat filters of the same radii as are used in the real universe, and
fit these to our Edgeworth model exactly as was done above.  We wish
to compare our results to the predictions of perturbation theory, and
therefore work in real space, not in redshift space.
 
For Gaussian filters, $N$-body simulation checks for $S_3$ and $S_4$
(Juszkiewicz \etal\ 1993; Bernardeau 1994b; {\L}okas \etal\ 1995) have
successfully matched the predicted values of $S_3$ and $S_4$ for the
$n = -1$ power spectrum.  The counts in cells method uses a top hat
filter, for which perturbation theory predicts $S_3^p = 2.86$ and
$S_4^p = 13.89$ for the $n=-1$ power spectrum of our simulations, from
equation~(\ref{eq:s3perturb})
and~(\ref{eq:s4perturb})\footnote{Actually, the values of $S_3^p$ and
$S_4^p$ we quote are for $\Omega_0 = 1$.  For $\Omega_0 = 0.3$, $S_3^p
= 2.89$ (Bouchet \etal\ 1995), differing by only 0.03 from the value
quoted above.  We are not aware of a direct analytic calculation of
the $\Omega_0$ dependence of $S_4$, but Bernardeau (1994a) shows that
it is very insensitive to $\Omega_0$.}.  Unlike the real \iras\ data,
biasing is not an issue we have to consider in the $N$-body
simulations, and therefore we expect the perturbation theory
prediction to agree quantitatively with the results of our method.


\begin{figure}[p]
\plotfiddle{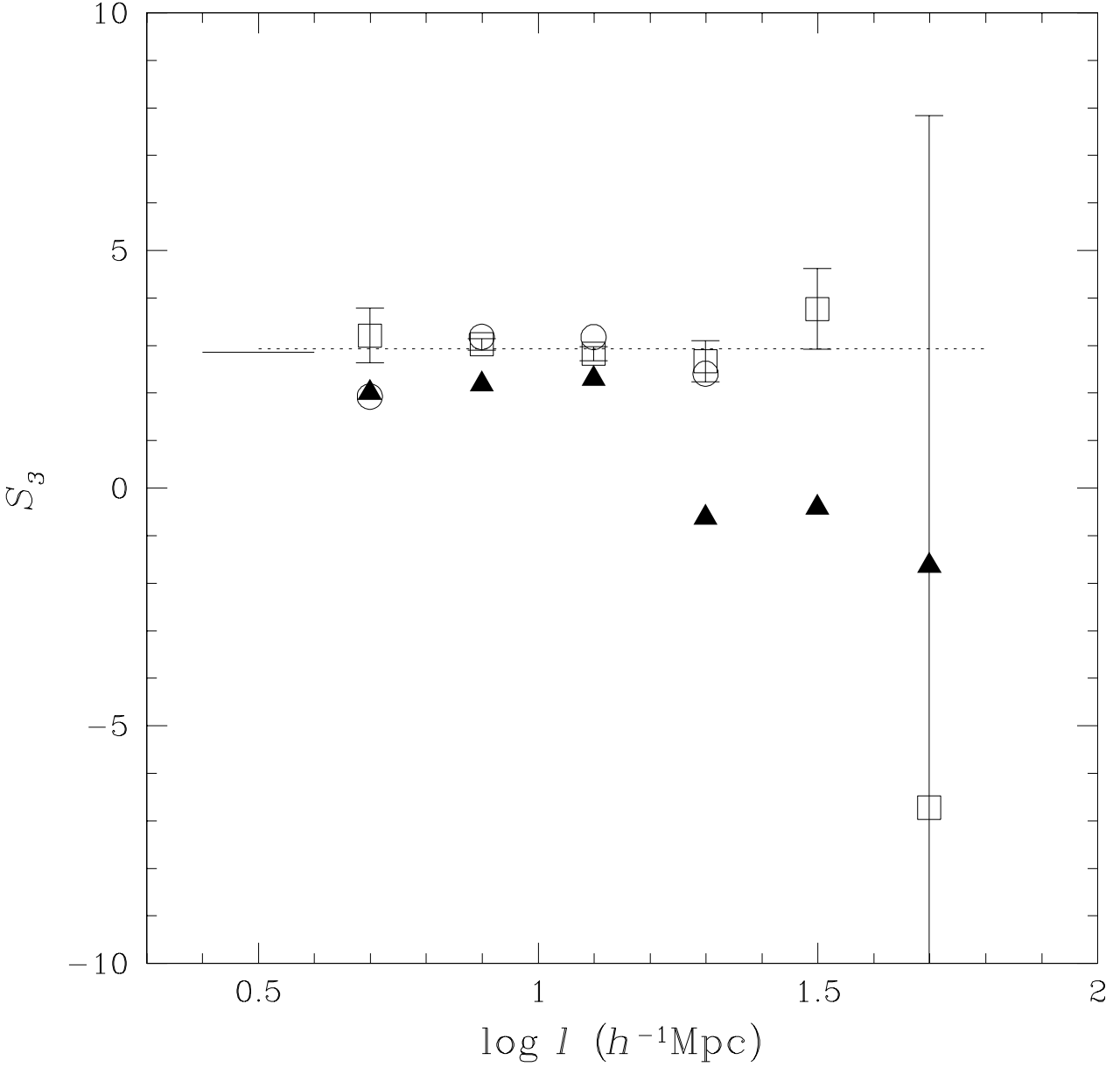}{6.3in}{0}{90}{90}{-270}{-140}
\caption{ 
Weighted average of $S_3$ vs scale, of the Edgeworth approximation
applied to \iras\ mock catalogs (open squares). The average value
$\langle S_3 \rangle = 2.93 \pm 0.09 \ (1\,\sigma)$ is indicated by
the dotted line. The open circles are the results of the moments
method applied to an $N$-body simulation sampled at 0.01 Mpc$^{-3}$,
yielding an average $\langle S_3 \rangle = 2.90\pm 0.64 \
(1\,\sigma)$.  The solid line on the left is the perturbation theory
prediction $S_3^p = 2.86$ for $n=-1$.  All three methods agree
well. The triangles are the results for the moments method applied to
the sparsely sampled mock catalogs.  It breaks down at scales larger
than $l \sim 16\,h^{-1}$ Mpc due to finite-volume effects (which we do
not correct for here), and results in a biased value of $S_3$.
\label{fig:ls3nbody}
	}
\end{figure}


\begin{figure}[p]
\plotfiddle{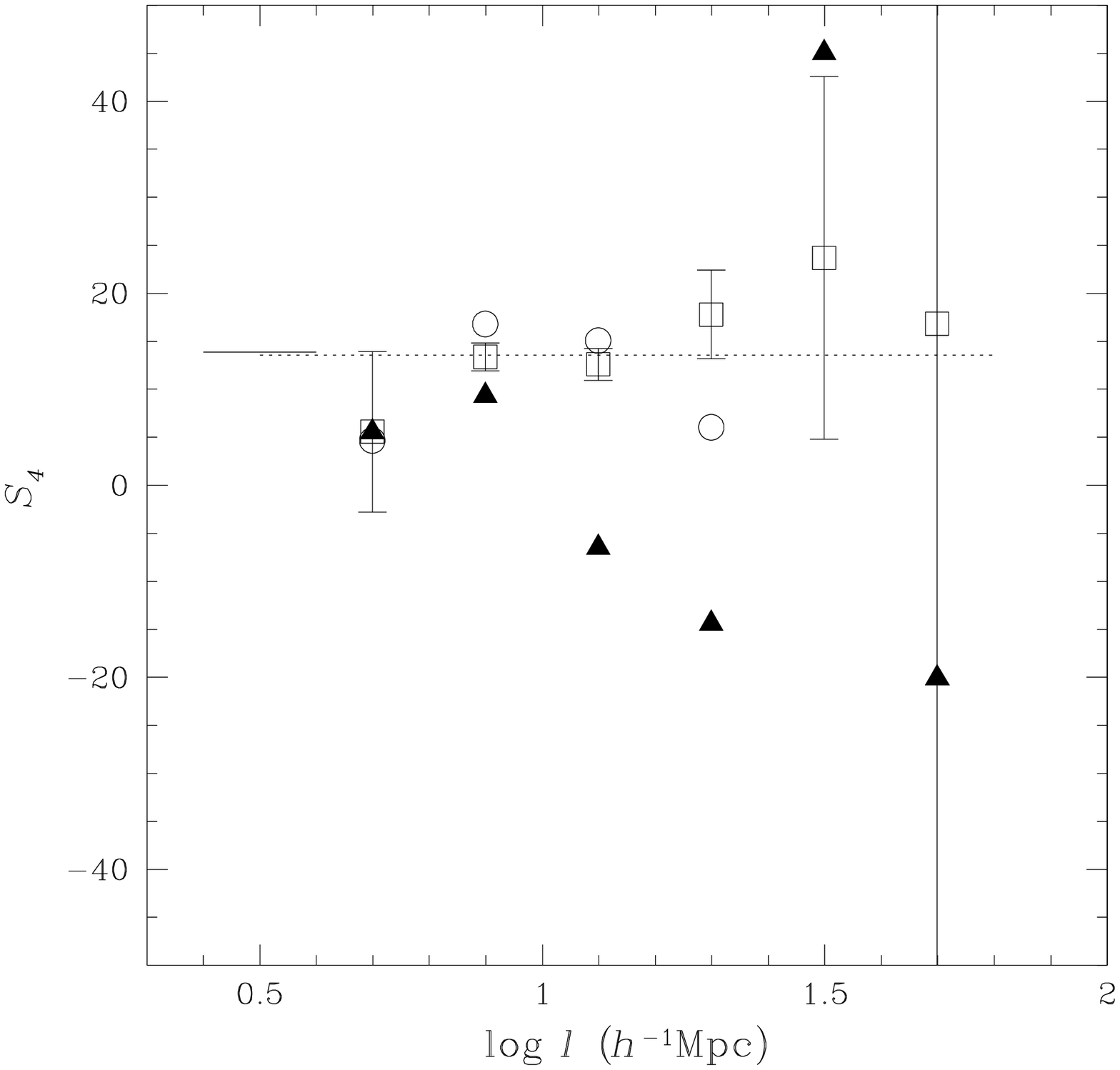}{6.9in}{0}{90}{90}{-270}{-108}
\caption{ 
$S_4$ vs.~scale for the mock catalogs. Open squares denote
the weighted average of the Edgeworth approximation applied to mock
catalogs, as in Fig~\ref{fig:ls3nbody}.  The average over scales is
$\langle S_4 \rangle = 13.54\pm 0.53$ (dotted line), in agreement with
the perturbation theory prediction $S_4^p = 13.89$ (solid line) and
that of the moments method applied to the full $N$-body simulation
(open circles). The moments method on sparsely sampled mock catalogs
(triangles) breaks down on scales larger than $l\sim 10$ \mpc.
\label{fig:ls4nbody}
	}
\end{figure}


Figure~\ref{fig:ls3nbody} shows the values obtained for $S_3$ as a
function of scale from the $N$-body model.  Open squares denote the
weighted average of the values obtained by the third-order Edgeworth
model to the CPDF's of the mock catalogs, at each scale.  The data
appear consistent with scale-invariance, as indeed one expects for
this power spectrum (cf., the discussion in Colombi, Bouchet, \&
Hernquist 1996).  Averaging over all scales gives $\langle S_3 \rangle
= 2.93 \pm 0.09$ (dotted line), in beautiful agreement with the
perturbation theory value (solid line).

The open circles indicate the results of applying the moments method
to the $N$-body simulation sampled at a density of $0.01\,
\mbox{Mpc}^{-3}$.  At this sampling density, the Poisson noise is
relatively small, and the moments method yields an equal-weight
average $\langle S_3 \rangle = 2.90 \pm 0.64$, remarkably similar to
the perturbation theory results, and very much consistent with the
Edgeworth approximation results of the mock catalogs.  Note that the
scatter around the mean of the determinations of $S_3$ using the
moments method from the densely sampled data is appreciably larger
than that of the Edgeworth expansion results based on the much sparser
mock catalogs.

When we apply the moments method to the more sparsely sampled mock
catalogs (triangles in Figure~\ref{fig:ls3nbody}), we find that the
values of $S_3$ are consistently biased low, an effect which worsens
at larger scales.  This is due to the finite-volume effect discussed
in \S~\ref{sec:intro}, and perhaps also finite-sampling effects as
well.  We have found that the CPDF for these sparse samples on large
scales never clearly reaches the asymptotic exponential tail
discussed, e.g., in Colombi \etal\ (1995), and thus we cannot fit the
tail to correct for these effects.

Figure~\ref{fig:ls4nbody} shows results for $S_4$; the symbols have
the same meaning as in Figure~\ref{fig:ls3nbody}. The average value of
$S_4$ obtained from the mock catalogs via the Edgeworth approximation
is $\langle S_4 \rangle = 13.54\pm 0.53$ (open squares and dotted
line), consistent with the perturbation theory prediction of $S_4 =
13.89$ (solid line).  The moments method applied to the densely
sampled $N$-body points (open circles) gives $\langle S_4 \rangle =
12.82 \pm 6.62$, which is very similar to the Edgeworth approximation
results (open squares), albeit again with larger errors, whereas the
moments method results from the mock catalog (triangles) begin to
break down at even very small scales.  We conclude that our method
gives unbiased estimates of $S_3$ and $S_4$, while the moments method
is systematically biased low for sparse samples by finite-volume
effects, which cannot easily be corrected for.

The last test we perform on $N$-body simulations is to check the
estimation of our error bars, via the method explained in
\S~\ref{sec:errors}.  For several specific values of sample size $R$
and smoothing scale $l$, we draw a series of 50 mock catalogs randomly
from the simulation, compute $\tilde P(N)$ for each, and fit them to
find $S_3$ and $S_4$, and their estimated errors, in each case.  We
found in every case that the mean of the estimated errors agreed with
the standard deviation of the individual values of $S_3$ and $S_4$ to
within 10\%, implying that the error estimates are correct in the
mean.  Of course, this only works when we scale ${\cal L}$ by the
ratio of $\chi^2$ to the number of degrees of freedom, as discussed in
\S~\ref{sec:errors}; if we do not do this, our errors are
overestimated by factors of 2 or 3.  However, we emphasize here that
our error estimation is not done in a rigorous way; our approach in
\S~\ref{sec:errors} is empirical at best, and further tests of our
errors with simulations with a variety of power spectra are needed to
justify these errors fully.

\section{Results for \iras\ Galaxies}
\label{sec:results}

\begin{figure}[p]
\plotfiddle{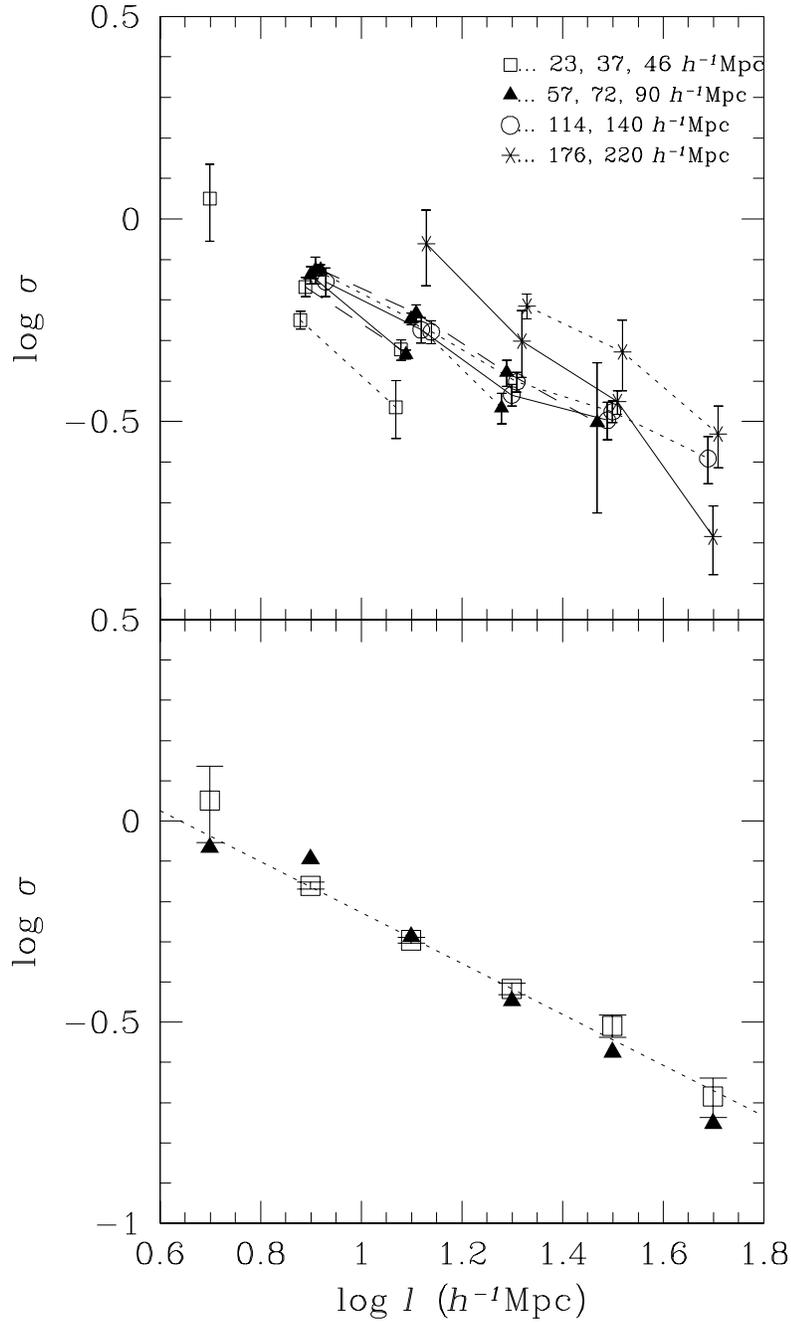}{5.9in}{0}{100}{100}{-310}{-160}
\caption{
A log-log plot of determinations of $\sigma(l)\ \mbox{vs}\ l$ from the
third order Edgeworth expansion fit. In the upper panel, points from
each subsample are connected; different symbols denote different
sample sizes as shown in the legend. Each symbol is used for several
samples: the solid line connects the points for the smallest sample,
the dotted line the next, and the dashed line the largest sample that
the symbol denotes.  The points are slightly staggered to show the
error bars.  The lower panel shows squares as weighted averages of the
top panel.  The dotted line is a least square fit to the squares, and
the triangles in the lower panel are an equal weighted average of the
results from the moments method.
\label{fig:lsigma}   
	}
\end{figure}

Table 2 shows that the best-fit values of $\sigma$, $S_3$ and $S_4$
are often considerably different from those found by the moments
method.  Figure~\ref{fig:lsigma} shows $\log \sigma \equiv 1/2 \log
\bar {\xi}_2$ versus the smoothing scale $l$, for the values found in
the third order Edgeworth expansion fit. The upper panel shows all
cases separately, with different symbols representing different sample
sizes and lines connecting the points found within each subsample.  As
the size $R$ of the subsample increases, $\sigma (l)$ grows for any
given $l$; more luminous galaxies show stronger clustering than do
less luminous galaxies (B93). This luminosity effect agrees well with
that of the moments method quantified in B93. However, within each
subsample the values trace a power law reasonably well, as does the
weighted average given by the squares in the lower panel.  The dotted
line in the lower panel is a least square fit to a power law.  The
average of the slopes at each value of $R$ is $\gamma/2 = 0.87\pm
0.08$, crossing unity at an average value of $l_0 = 5.07\pm 1.45
\,h^{-1}\mbox{Mpc}$, while the moments method (triangles in the lower
panel) yields $\gamma/2 = 0.80 \pm 0.03, \ l_0 = 5.44\pm
0.53\,h^{-1}\mbox{Mpc}$ (B93). This is quite reassuring, considering
the sensitivity of $\sigma$ to the order of the fit (Table~2).  The
large error bar in $l_0$ is due to its sensitivity on luminosity.


\begin{figure}[p]
\plotfiddle{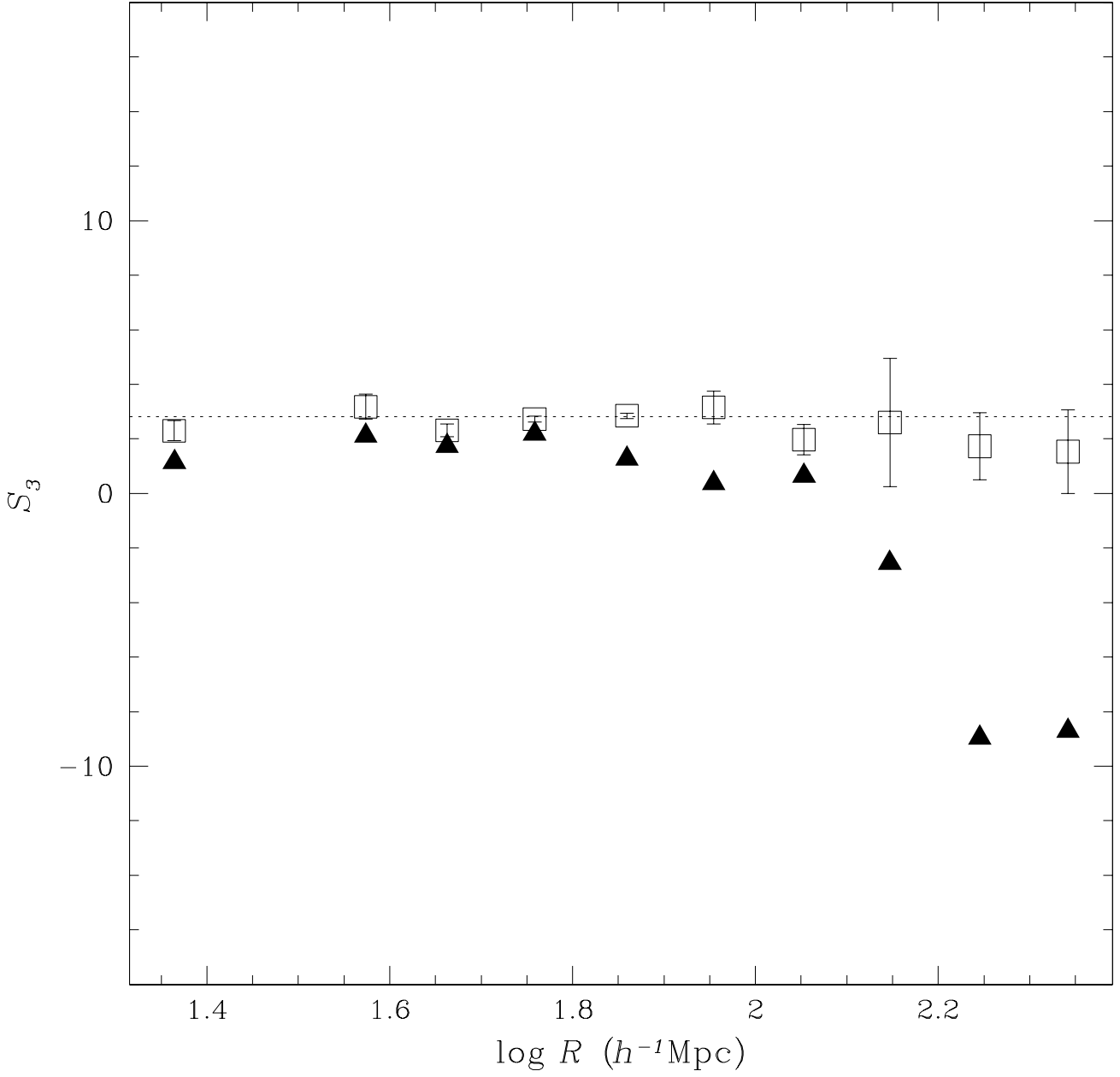}{6.9in}{0}{90}{90}{-270}{-108}
\caption{ 
Weighted averages of $S_3$ results from the Edgeworth approximation as
a function of subsample size $R$ (open squares), from Table~3.  The
triangles are the results of the moments method.  The dotted line
denotes the global average of the Edgeworth approximation, $\langle
S_3 \rangle = 2.83 \pm 0.09$.
\label{fig:rs3}
	}
\end{figure}

\begin{figure}[p]
\plotfiddle{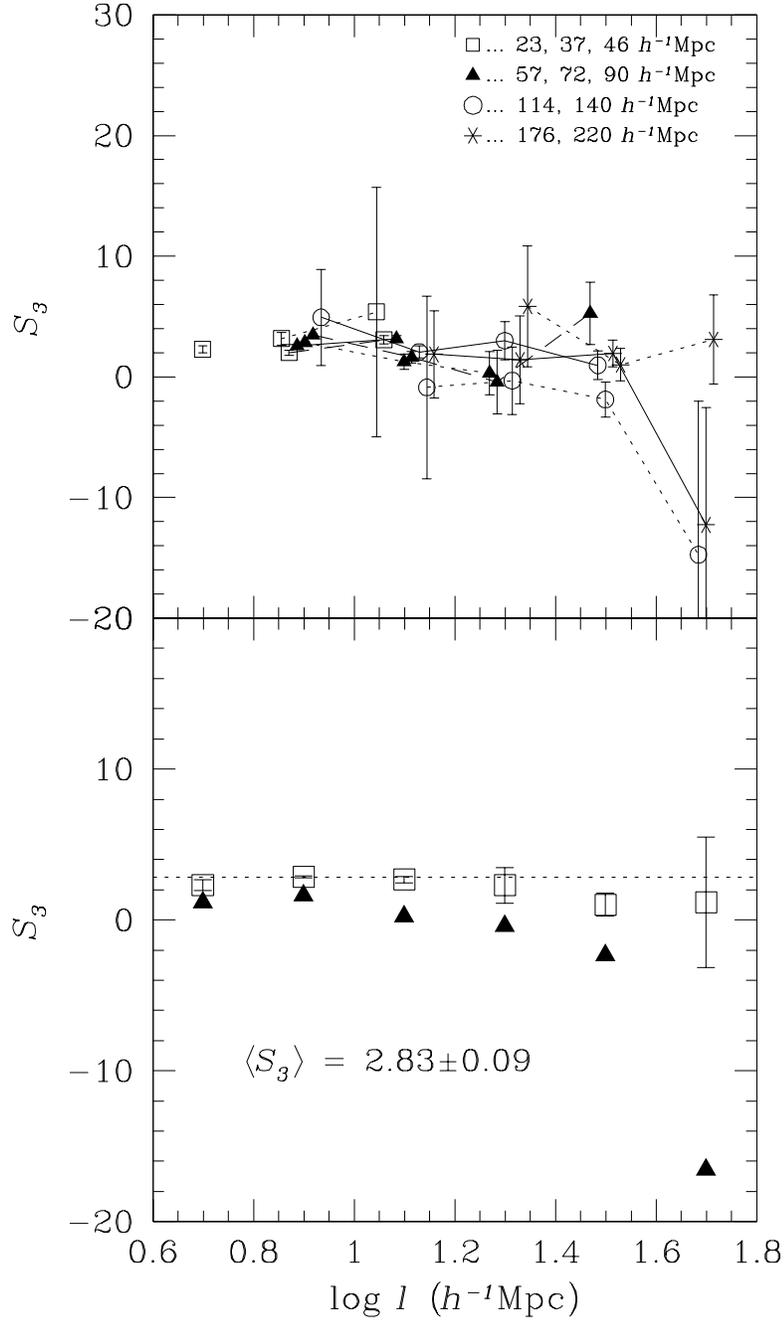}{6.2in}{0}{100}{100}{-310}{-160}
\caption{ 
All determinations of $S_3$ as a function of scale, on the top
panel. As in Figure~\ref{fig:lsigma}, different symbols indicate
different subsamples, with lines connecting the values within the same
subvolume. Bottom panel shows weighted averages of the Edgeworth
approximation results as open squares; the averages of the results of
the moments method are indicated by triangles.  The dashed line is the
global average of the Edgeworth approximation results, $\langle S_3
\rangle = 2.83\pm 0.09$.
\label{fig:ls3}
	}
\end{figure}

\begin{figure}[p]
\plotfiddle{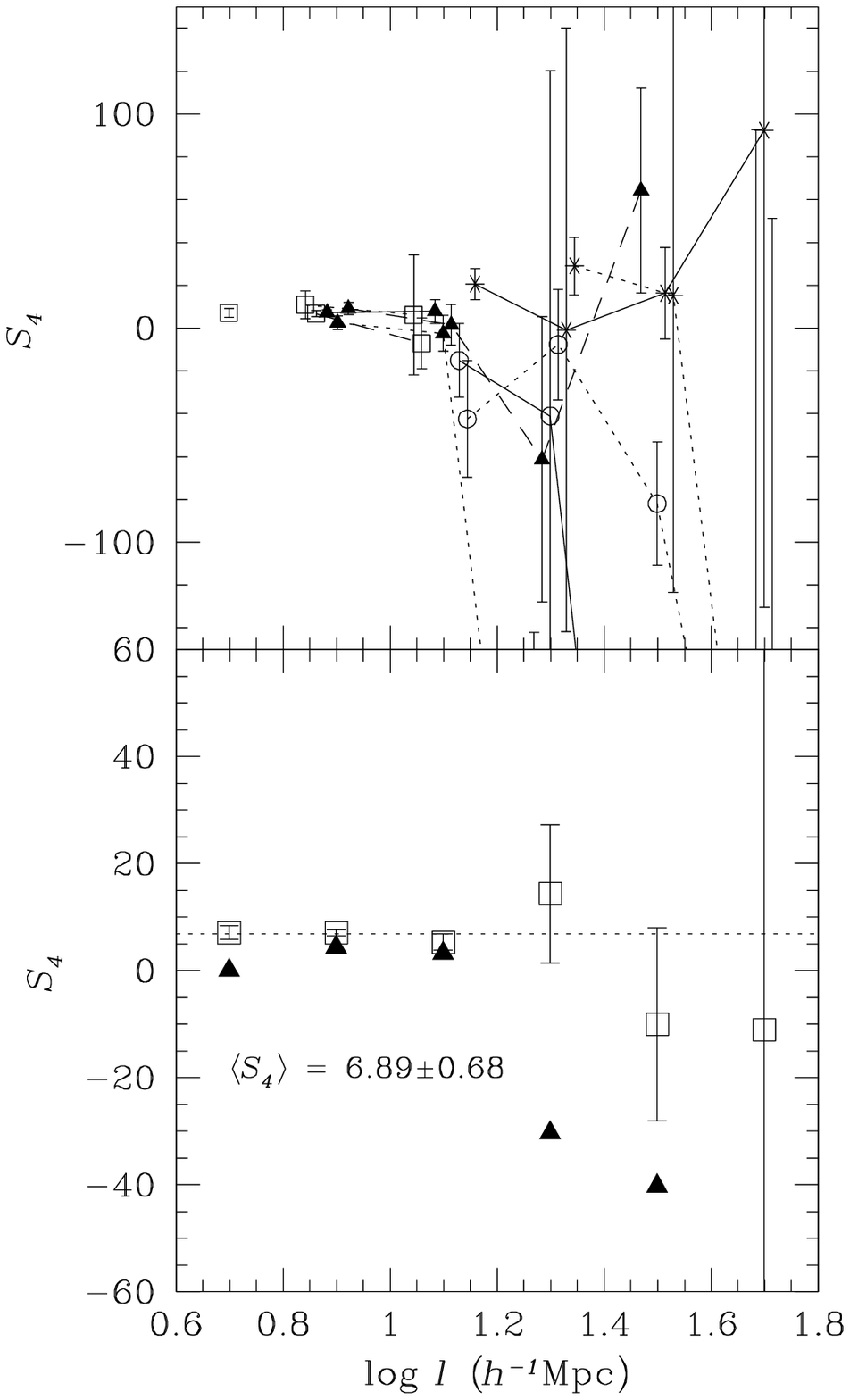}{6.9in}{0}{100}{100}{-310}{-140}
\caption{ $S_4$ as a function of scale. The symbols have the same
meaning as Figure~\ref{fig:ls3}.
\label{fig:ls4}
	}
\end{figure}

The weighted average values of $S_3$ and $S_4$ found by fitting the
third order Edgeworth expansion are summarized in Tables~3 (averaging
at a given sample size $R$) and 4 (averaging at a given scale $l$),
and are plotted in Figures~\ref{fig:rs3}, \ref{fig:ls3}, and
\ref{fig:ls4}.  Note that our likelihood method gives meaningful error
bars on the values of $S_3$ and $S_4$, enabling us to perform a
weighted average of our determinations on different scales and from
different subsamples, assuming that they are independent of
scale\footnote{There is some covariance due to the fact that the same
galaxies are used for the determination of the parameters on different
scales; we ignore this effect here.}.


The figures also show results for the moments method.  They are
systematically lower than the Edgeworth results, due to finite-volume
effects.  As with the $N$-body tests described in the previous
subsection, we found that on these large scales, the CPDF never
reached the asymptotic form that would allow us to correct for these
effects (compare with Fry \& Gazta\~naga 1994, who did perform this
correction, but only on scales below 10$\,h^{-1}$Mpc).  Error bars for
values determined from the moments method can be calculated (Szapudi
\& Colombi 1996), but are quite complicated. Following B93, we use the
standard deviation of determinations of $S_3$ and $S_4$ as our error
bar for the moments method, including only positive values.

The Edgeworth approximation yields values of $S_3$ and $S_4$ (open
squares in Figures~\ref{fig:rs3}, \ref{fig:ls3}, and \ref{fig:ls4})
which show no statistically significant dependence on $R$ or $l$,
consistent with the scale-invariant hypothesis (cf., discussion
below). Without finite-volume corrections, this scale-invariance is
not apparent from the moments results.  This is not an issue with the
Edgeworth method; it is quite insensitive to the tail of the CPDF, and
thus no correction for finite-volume effects is needed.

We find a global average of $\langle S_3 \rangle = 2.83 \pm
0.06$\footnote{ This error bar is smaller than our final quoted value
of 0.09 ; see discussion below.}, which is appreciably larger than the
value $\langle S_3 \rangle = 1.5 \pm 0.5$ found in B93.  This is
consistent with the bias in the moments method we found in
\S~\ref{sec:nbodytest}; we showed there that the Edgeworth method is
unbiased, and we therefore believe that our determination of $\langle
S_3 \rangle$ for \iras\ galaxies supersedes that of B93.  Our results
are in better agreement with the moments methods results for \iras\
from Fry \& Gazta\~naga (1994), which found $S_3 =2.1 \pm 0.3, S_4 =
7.5 \pm 2.1$, in the range 3-10$\,h^{-1}$Mpc, and Meiksin \etal\
(1992) from the angular distribution of \iras\ galaxies ($S_3 = 2.2\pm
0.2 , S_4 = 10\pm 3$ in the range 4-10$\,h^{-1}$Mpc)\footnote{ Note,
however, that one does not expect the angular and spatial $S_N$ to be
identical (cf., Bernardeau 1995).}.

Our result uses the third-order Edgeworth expansion; if
we use the second-order expansion, we find $\langle S_3 \rangle = 2.80
\pm 0.11$, which is consistent with the value from the third order
fit.   In both cases, the average is dominated by the point with the
smallest errors, $\langle S_3 \rangle = 2.87 \pm 0.06$ at
8$\,h^{-1}$Mpc. 

Given the estimated errors on $S_3$ for each value of $R$ and $l$, we
can do an {\it a posteriori\/} test of the hypothesis that $S_3$ is
independent of sample, by computing the $\chi^2$-like statistic:
\begin{equation} 
{\cal R} = \sum_{{\rm realizations}\ i}{{(S_{3,i} - \langle S_3
\rangle) ^2} \over {\sigma_{S3,i}^2}},
\label{eq:s3chi2} 
\end{equation}
where the sum is over all different values of $R$ and $l$.  We find
that $\cal R$ exceeds the number of measurements of $S_3$ by a factor
of two. This is due to several effects: cosmic variance, the
approximation we have made that the errors in $S_3$ determined at
different scales from the same data are statistically independent, and
possible real scale-dependence of $S_3$ caused by higher-order effects
(cf., Colombi \etal\ 1996).  Our final error bar should reflect this
increased scatter, and so we multiply the error on $\langle S_3
\rangle$ by ${\cal R}^{1/2} = 1.4$, yielding $\langle S_3 \rangle =
2.83 \pm 0.09$.  It would be interesting to separate out these effects
with a $n = -2$ simulation, in which these higher-order effects should
be more important (Colombi \etal\ 1996); this is work for the future.

Figure~\ref{fig:ls4} shows the determination of $S_4$ as a function of
scale.  For scales larger than $\log l = 1.3$, $S_4$ becomes very
uncertain, with large error bars.  However, the weighted average
values of $S_4$ in the bottom panel stay close to the total average
value of $\langle S_4 \rangle = 6.89 \pm 0.48$, with error bars
overlapping that value at all scales.  We carried out the $\chi^2$
test of equation~(\ref{eq:s3chi2}) for $S_4$ as well, finding ${\cal
R} = 2$ again.  We thus also multiply our quoted error on $\langle S_4
\rangle $ by 1.4, yielding $\langle S_4 \rangle = 6.98 \pm 0.68$.
This again is larger than, albeit statistically consistent with, the
moments method results of $\langle S_4 \rangle = 4.4 \pm 3.7$ (B93),
and is in agreement with Meiksin \etal\ (1992) and Fry \& Gazta\~naga
(1994), quoted above.

Gazta\~naga (1992) used the moments method to find $\langle S_3
\rangle = 1.94\pm 0.07$ for optically selected galaxies.  Gazta\~naga
(1992) and B93 claim that the lower value of $S_3$ for \iras\ galaxies
can be attributed to the underrepresentation of \iras\ galaxies in
dense cluster cores (cf., Strauss \etal\ 1992).  As a test of this,
B93 gave extra weight to the \iras\ clusters to match the
overdensities seen in optically selected galaxies; they found the
$S_N$ to be quite sensitive to this: $\langle S_3^b \rangle = 3.71 \pm
0.95$ and $\langle S_4^b \rangle = 23.6 \pm 12.1$.  This demonstrates
the high sensitivity of the moments method to dense clusters; the
moments are heavily weighted by the tail of the CPDF, and this effect
is even more important for $S_4$ than for $S_3$. This sensitivity is
very dangerous, given the fact that the tail is generally hard to
measure with accuracy, as we have seen.  We have fit the Edgeworth
model to the CPDF of these cluster-boosted counts, and found $\langle
S_3^b\rangle = 2.65\pm 0.09$, $\langle S_4^b \rangle = 7.79 \pm 0.67$,
within 2 standard deviations of the unboosted results above.  This is
as expected; boosting the clusters only affects the CPDF in the tail,
and therefore this has only a small effect on our fits.

One might argue that this result is misleading; if our Edgeworth model
is a good fit to the full CPDF before the cluster-boosting, it {\it
cannot\/} be a good fit afterwards, because the tail has changed
dramatically, even though it had very little effect on the rest of the
CPDF.  Clearly, the moments method and our method cannot agree in both
cases, independent of issues of finite volume effects.  With our
method, $S_3$ and $S_4$ are determined from a fit to that part of the
CPDF that is close to mean density, and therefore is not highly
non-linear, while the moments method is quite sensitive to the
non-linear tail.  Thus our method can measure $S_3$ and $S_4$ in the
weakly nonlinear regime, even when strong clustering is present.

The effective power-law index for \iras\ galaxies is $n = -1.4$
(Fisher \etal\ 1993), which would predict that $S_3 = 3.26$ and $S_4 =
22.4$.  Why do our results differ from these values?  If the \iras\
galaxy distribution were biased with respect to the underlying mass,
one would expect that the skewness and kurtosis would be
systematically affected.  The linear bias model, $\delta_g =
b\delta_M$, where $\delta_g$ is the observed galaxy density field and
$\delta_M$ is the underlying mass density contrast, predicts $S_{3,g}
= S_{3,M} / b$ and $S_{4,g} = S_{4,M}/b^2$.  However, as Fry \&
Gazta\~naga (1993) point out, we cannot consider high-order
correlations without also considering the possibility of high-order
bias (Gazta\~naga \& Frieman 1994; Juszkiewicz \etal\ 1995):
\begin{equation}
	\delta_g = f(\delta) = \sum_{k=1}^{\infty} {b_k \over k!}
			\delta_M^k  .
\end{equation}
This leads to:
\begin{eqnarray}
	 S_{3g} &=& S_{3M} /b_1 + 3b_2 / b_1^2 + 
			{\cal O}( \langle \delta^2\rangle )\\
	 S_{4g} &=& S_{4M} /b^2 + 12S_{3M} b_2 / b_1^3 + 4b_3 / b_1^3 
		    + 12 b_2^2 / b_1^4  + {\cal O}(\langle \delta^2\rangle ) .
\end{eqnarray}
Thus without external information on the detailed form of the biasing
relation, we cannot make a direct comparison of our results with those
from perturbation theory.

\section{Conclusions}

We have measured the count probability distribution function via the
counts-in-cells method for 10 volume limited subsamples of the \iras\
1.2 Jy redshift survey, exactly as in B93. There are various
approaches to measure the skewness $S_3$ and kurtosis $S_4$ of the
probability distribution function of the underlying density field.
B93 calculated these quantities for this sample using the moments
method. They found scale invariance of $S_3$ to $l\sim 25$ \mpc, with
an average value of $S_3 = 1.5 \pm 0.5$.  However, the moments method
is very sensitive to the high density tail of the CPDF, making the
values of $S_3$ and $S_4$ sensitive to finite-volume and
finite-sampling effects.  These effects can be corrected for (Fry \&
Gazta\~naga 1994; Colombi \etal\ 1995; Szapudi \& Colombi 1996;
Szapudi \etal\ 1996), at least in the strongly non-linear regime with
dense sampling.  We work here with a very sparse redshift survey in
the weakly non-linear region ($\bar{\xi_2}<1$), and find that the CPDF
never properly reaches the asymptotic limit that allows one to correct
for finite-volume effects.  We propose a method less sensitive to the
tails of the CPDF, a maximum likelihood fit of the Edgeworth expansion
convolved with a Poissonian, to the observed CPDF.  The Edgeworth
expansion is valid only in the weakly nonlinear regime ($\sigma < 1$);
unlike the moments method, it cannot be applied on very small scales.

We have tested our method with \iras\ mock catalogs extracted from
$N$-body simulations; we find that the derived values of $S_3$ and
$S_4$ are consistent with the analytic predictions from perturbation
theory, as well as from the moments method as derived from densely
sampled $N$-body points.  Moreover, our estimated errors are
consistent with the scatter in $S_3$ and $S_4$ seen in multiple
realizations of the sample.  The results from the moments method in
these sparse mock catalogs are systematically biased low, especially
on large scales, due to finite-volume effects.  Hence we conclude that
the Edgeworth approximation is much more reliable and robust than is
the moments method, especially in sparse samples and in the weakly
non-linear regime, where there is no simple method to correct for
these effects.

The resulting values of $S_3$ and $S_4$ are found to be $\langle S_3
\rangle = 2.83 \pm 0.09$ and $\langle S_4 \rangle = 6.89 \pm 0.68$,
significantly higher than the results of B93, but consistent with
Meiksin \etal\ (1992) and Fry \& Gazta\~naga (1994).  These results
are quite insensitive to the fact that \iras\ galaxies are
underrepresented in cluster cores.  Both $S_3$ and $S_4$ are
independent of scale within the errors from 5 $h^{-1}$Mpc to 50
$h^{-1}$Mpc.

We have shown that the data are consistent with the scale-invariant
hypothesis.  It would be very interesting to compare these results
with those from various specific models with non-Gaussian initial
conditions, to see at what level we might be able to rule them out. 

Application of the Edgeworth approximation to optical samples should
be interesting, especially since previous work has shown discrepancies
with the \iras\ sample, attributed to the underrepresentation of
\iras\ galaxies in clusters.  Also interesting would be to apply this
technique to angular surveys such as the APM galaxy sample, where we
could carry this technique out to appreciably higher order.  We also
look forward to applying this technique on the spectroscopic and
photometric data of the Sloan Digital Sky Survey (cf., Gunn \&
Weinberg 1995), which will allow us to probe appreciably larger
spatial scales.

There is also further work to be done on the method itself.  Our
understanding of the errors and covariances in the CPDF, and therefore
the errors in our derived parameters, is poor, and thus our final
errors are not rigorously justified.  As we have seen in
\S~\ref{sec:results}, the $\chi^2$ test (equation~\ref{eq:s3chi2})
suggests that our method of obtaining error bars can hide interesting
higher order effects.  In particular, without good {\it a priori\/}
errors, we cannot do a proper test of goodness of fit of our model.
Further analytical work in this direction is needed, together with
more extensive tests with simulations over a wider range of
conditions.

\acknowledgements 
We thank R. Juszkiewicz and D. Weinberg for important discussions at
the outset of this project, E. Gazta\~naga, S. Colombi,
 and an anonymous referee
for useful comments, and D. Weinberg for the $N$-body simulations used
in this paper.  MAS gratefully acknowledges the support of an Alfred
P. Sloan Foundation Fellowship, as well as the support of NASA
Astrophysical Theory Grant NAG5-2882.  RSK acknowledges the support of
an Assistant in Research Tuition Award from Princeton University, and
NSF grant AST93-15368.

\clearpage

\begin{deluxetable}{l}
\tablecaption{Hermite Polynomials}
\tablecolumns{1}
\tablewidth{180pt}
\tablehead{}
\startdata
$H_0 (\nu) = 1$   \nl
$H_1 (\nu) = \nu$ \nl
$H_2 (\nu) = \nu^2 - 1$ \nl
$H_3 (\nu) = \nu^3 - 3\nu$ \nl
$H_4 (\nu) = \nu^4 - 6\nu^2 +3$ \nl
$H_5 (\nu) = \nu^5 - 10\nu^3 + 15\nu$ \nl
$H_6 (\nu) = \nu^6 - 15\nu^4 + 45\nu^2 - 15$ 
\enddata
\label{table:1}
\end{deluxetable}
\clearpage

\begin{deluxetable}{cccclllc}
\tablecaption{Results of representative fits }
\tablecolumns{8}
\tablehead{
	   \colhead{Max. Radius\tablenotemark{a}} & 
	   \colhead{$l\ $\tablenotemark{b}} & 
	   \colhead{$\bar{N}\ $\tablenotemark{c}} & \colhead{order} &
           \colhead{$\sigma(l)$} & \colhead{$S_3 (l)$} & 
	   \colhead{$S_4 (l)$} & \colhead{${\cal L}\ $\tablenotemark{d}}
	  }

\startdata
\tablenotemark{*}
$37.46$ & $7.92$ & $6.57$ & m \tablenotemark{e} &  $0.70$	 & $1.97$	  & $3.22$	    & $384.49$ \nl \cline{5-8}
	&	 &	  & $1$ & $0.61\pm 0.09$ & ...            & ...  	    & $418.27$ \nl 
	&	 & 	  & $2$ & $0.56\pm 0.04$ & $4.15\pm 0.83$ & ...	 	    & $369.02$ \nl
	&	 &	  & $3$ & $0.56\pm 0.05$ & $3.17\pm 0.66$ & $10.82\pm 4.08$ & $363.01$ \nl
\hline
\tablenotemark{*}
$46.01$ & $7.92$ & $3.54$ &  m  & $0.73$	 & $1.74$	  & $2.69$	    & $895.01$ \nl \cline{5-8}
	&	 &	  & $1$ & $0.68\pm 0.08$ & ...   	  & ...		    & $1050.8$ \nl
	&  	 &	  & $2$ & $0.68\pm 0.04$ & $2.70\pm 0.44$ & ...		    & $889.81$ \nl
	&	 &	  & $3$ & $0.78\pm 0.10$ & $2.00\pm 0.28$ & $6.80\pm 0.93$  & $882.27$ \nl 
\hline
$57.34$ & $7.92$ & $2.28$ &  m  & $0.85$	 & $2.13$	  & $6.49$	    & $1187.2$ \nl \cline{5-8}
	&	 &	  & $1$ & $0.74\pm 0.06$ & ...	   	  & ...		    & $1846.9$ \nl
	&	 &	  & $2$ & $0.73\pm 0.05$ & $3.43\pm 0.56$ & ...		    & $1187.1$ \nl
	&   	 &	  & $3$ & $0.77\pm 0.02$ & $2.56\pm 0.11$ & $7.38\pm 0.76$  & $1170.6$ \nl
\hline
\tablenotemark{*}
$57.34$ & $12.56$& $8.39$ &  m  & $0.53$	 & $2.26$ 	  & $4.90$	    & $2066.3$ \nl \cline{5-8}
	& 	 &	  & $1$	& $0.48\pm 0.08$ & ...		  & ...		    & $2163.1$ \nl
	&	 &	  & $2$	& $0.46\pm 0.02$ & $3.53\pm 0.57$ & ...		    & $2034.0$ \nl
	&	 &	  & $3$	& $0.45\pm 0.02$ & $3.15\pm 0.32$ & $7.83\pm 3.46$  & $2033.9$ \nl
\hline
$72.35$ & $7.92$ & $0.98$ &  m	& $0.81$	 & $1.30$	  & $1.12$	    & $701.67$ \nl \cline{5-8}
	&	 &	  & $1$	& $0.67\pm 0.06$ & ...	 	  & ...		    & $1005.0$ \nl
	&	 &	  & $2$	& $0.74\pm 0.06$ & $3.64\pm 0.79$ & ...		    & $527.17$ \nl
	&	 &	  & $3$	& $0.73\pm 0.02$ & $2.85\pm 0.11$ & $2.51\pm 1.94$  & $499.30$ \nl
\hline
$72.35$	& $12.56$& $3.82$ &  m	& $0.54$	 & $1.32$	  & $4.35$	    & $140.29$ \nl \cline{5-8}
	&	 &	  & $1$ & $0.57\pm 0.08$ & ...		  & ...		    & $142.70$ \nl
	&	 & 	  & $2$ & $0.57\pm 0.02$ & $1.39\pm 0.46$ & ...		    & $138.81$ \nl
	& 	 & 	  & $3$	& $0.52\pm 0.12$ & $1.28\pm 0.80$ & $-2.4 \pm 14.5$ & $138.63$ \nl
\hline 
$90.02$ & $7.92$ & $0.45$ &  m	& $0.87$	 & $1.63$	  & $1.34$	    & $1493.5$ \nl \cline{5-8}
	& 	 &	  & $1$ & $0.58\pm 0.07$ & ...		  & ...		    & $1635.6$ \nl
	&	 &	  & $2$ & $0.74\pm 0.06$ & $4.69\pm 0.88$ & ...		    & $935.40$ \nl
	&	 &	  & $3$ & $0.76\pm 0.11$ & $3.45\pm 0.66$ & $9.17\pm 1.70$  & $929.90$ \nl
\hline
$90.02$	& $12.56$& $1.74$ &  m	& $0.57$	 & $0.86$	  &$-3.6$	    & $253.78$ \nl \cline{5-8}
	&	 &	  & $1$ & $0.59\pm 0.08$ & ...		  & ...		    & $268.85$ \nl 
	&	 &	  & $2$ & $0.59\pm 0.02$ & $2.01\pm 0.56$ & ...		    & $250.76$ \nl
	&	 &	  & $3$ & $0.58\pm 0.10$ & $1.61\pm 0.73$ & $1.44\pm 10.4$  & $250.87$ \nl
\hline
$113.02$& $19.71$& $3.17$ &  m	& $0.37$	 & $0.77$	  & $-21.6$	    & $357.50$ \nl \cline{5-8}
	&	 &	  & $1$ & $0.37\pm 0.07$ & ...		  & ...		    & $353.54$ \nl 
	&	 &	  & $2$ & $0.37\pm 0.03$ & $3.40\pm 2.49$ & ...		    & $348.66$ \nl
	&	 &	  & $3$ & $0.33\pm 0.06$ & $3.00\pm 1.96$ & $-41.36\pm 100.8$&$347.29$ \nl
\hline
\tablenotemark{*}
$175.99$& $31.54$& $1.78$ &  m	& $0.34$	 & $-0.38$	  & $-39.66$	    & $551.92$ \nl \cline{5-8}
	&	 &	  & $1$ & $0.35\pm 0.12$ & ...	 	  & ...		    & $546.20$ \nl
	&	 &	  & $2$ & $0.35\pm 0.03$ & $1.77\pm 0.94$ & ...		    & $545.30$ \nl
	&	 &	  & $3$	& $0.40\pm 0.16$ & $1.93\pm 1.40$ & $16.29\pm 13.6$ & $544.77$ \nl

\enddata
\tablenotetext{*}{Examples shown in Figure 2}
\tablenotetext{a}{Radius of subsamples ($h^{-1}$Mpc)}
\tablenotetext{b}{Radius of the sphere ($h^{-1}$Mpc)}
\tablenotetext{c}{Average number of points in the sphere of size $l$}
\tablenotetext{d}{${\cal L} = -2 \ln L$, $L$ is likelihood; values
shown here are scaled to the $\chi^2$ per degree of freedom (see
\S~\ref{sec:errors})}
\tablenotetext{e}{Initial values of the fit, obtained by moments method}
\label{table:2}
\end{deluxetable}

\begin{deluxetable}{crrrr}
\tablecaption{$S_3$ and $S_4$ for different subsamples}
\tablecolumns{5}
\tablehead{\multicolumn{1}{c}{Max. Radius ($h^{-1}$Mpc)} & \multicolumn{2}{c}{$S_3$} 
		& \multicolumn{2}{c}{$S_4$} \nl
	   \cline{2-3} \cline{4-5} 
           & \colhead{Moments\tablenotemark{a}} &
           \colhead{Edgeworth\tablenotemark{b}} & \colhead{Moments} &
           \colhead{Edgeworth} 
          }
\startdata
23.14 &1.15\hspace{.4in}  &$2.30\pm 0.37$& 0.12\hspace{.4in}  & $7.09 \pm 1.25$ \nl
37.47 & $2.11\pm 0.14$ & $3.18\pm 0.46$ & $10.6 \pm 3.87$ & $10.8\pm 9.62$ \nl
46.01 & $1.74\pm 0.01$ & $2.31\pm 0.24$ & $1.69 \pm 1.00$ & $6.59 \pm 0.93$ \nl
57.34 & $2.19\pm 0.06$ & $2.63\pm 0.11$ & $5.70 \pm 0.80$ & $7.39 \pm 0.75$ \nl
72.35 & $1.26\pm 0.06$ & $2.85\pm 0.10$ & $0.86 \pm 2.96$ & $2.35 \pm 1.92$ \nl
90.02 & $0.38\pm 0.93$ & $3.15\pm 0.13$ & $-14.7\pm 17.2$ & $8.10 \pm 1.66$ \nl
113.0 & $0.64\pm 1.00$ & $1.98\pm 0.67$ & $-12.2\pm 25.1$ & $-16.9\pm 10.6$ \nl
140.3 &$-2.55\pm 1.57$ &$ 2.60\pm 2.36$ & $-30.5\pm 6.75$ & $-7.61\pm 14.1$ \nl
176.0 &$-8.96\pm 10.8$ & $1.73\pm 1.23$ & $-54.2\pm 58.3$ & $17.4\pm 11.8$ \nl
219.7 &$-8.71\pm 6.91$ & $1.53\pm 1.53$ & $-167.\pm 178.$ & $14.5\pm 43.1$ \nl
\enddata
\tablenotetext{a}{Moments method : Equal weighted average. Errors $1\sigma$ dispersion}
\tablenotetext{b}{Results of Edgeworth expansion fit. Weighted average} 
\label{table:3}
\end{deluxetable}

\begin{deluxetable}{crrrr}
\tablecaption{$S_3$ and $S_4$ as a function of scale }
\tablecolumns{5}
\tablehead{\multicolumn{1}{c}{$l$ ($h^{-1}$Mpc)} & \multicolumn{2}{c}{$S_3$} 
		& \multicolumn{2}{c}{$S_4$} \nl
	   \cline{2-3} \cline{4-5}
           & \colhead{Moments\tablenotemark{a}} &
           \colhead{Edgeworth\tablenotemark{b}} & \colhead{Moments} &
           \colhead{Edgeworth} 
          }
\startdata
5.00 & 1.15\hspace{.41in} & $2.30 \pm 0.37$ & 0.12\hspace{.4in}  & $7.09 \pm 1.25$\nl
7.92 & $1.64 \pm 0.37$ & $2.87 \pm 0.06$ & $4.36 \pm 3.57$ & $7.06 \pm 0.56$\nl
12.6 & $0.21 \pm 2.78$ & $2.67 \pm 0.21$ & $3.18 \pm 12.7$ & $5.29 \pm 1.53$\nl
19.9 & $-0.39\pm 1.15$ & $2.30 \pm 1.19$ & $-30.5\pm 21.8$ & $14.6 \pm 12.9$\nl
31.5 & $-2.35\pm 3.11$ & $1.02 \pm 0.76$ & $-40.2\pm 5.92$ & $-10.0 \pm 18.0$\nl
50.0 & $-16.6\pm 9.02$ & $1.17 \pm 4.32$ & $-198.\pm 161.$ & $-11.0\pm 122.$\nl
\hline 
All  & $-1.76\pm 6.27$\tablenotemark{c} & $2.83 \pm 0.06$\tablenotemark{d} &
$-31.9\pm 79.0$\tablenotemark{c}		& $6.89 \pm 0.48$\tablenotemark{d} \nl
     & $1.54\pm 0.47$\tablenotemark{e}  &  & $5.8\pm 7.09$\tablenotemark{e}
\enddata

\tablenotetext{a}{Moments method : Equal weighted average. Errors $1\sigma$ dispersion}
\tablenotetext{b}{Results of Edgeworth expansion fit. Weighted average} 
\tablenotetext{c}{Equal weighted average of all values}
\tablenotetext{d}{Weighted average of all values. Note that our final
quotes are $\langle S_3\rangle = 2.83\pm 0.09$ and $\langle S_4
\rangle=6.89 \pm 0.68$ (see discussion in \S~\ref{sec:results}). }
\tablenotetext{e}{Equal weighted average of all positive values}
\label{table:4}
\end{deluxetable}

\clearpage

\clearpage

\begin{thebibliography}{}

\bibitem[Amendola 1994]{1}
Amendola, L. 1994, \apj, 430, L9

\bibitem[]{2}
Balian, R., \& Schaeffer, R. 1988, \apj, 335, L43

\bibitem[]{3}
Balian, R., \& Schaeffer, R. 1989, \aap, 220, 1

\bibitem[]{4}
Bernardeau, F. 1992, \apj, 392, 1

\bibitem[Bernardeau 1994]{5}
Bernardeau, F. 1994a, \apj, 433, 1

\bibitem[]{6} 
Bernardeau, F. 1994b, A\&A, 291, 697

\bibitem[]{7} 
Bernardeau, F. 1995, A\&A, 301, 309

\bibitem[Bernardeau, Kofman 1995]{BK}
Bernardeau, F. \& Kofman, L. 1995, \apj, 443, 479

\bibitem[]{8} Borgani, S. 1995, Phys. ~Rep., 251, 1

\bibitem[]{9} Bouchet, F. R., Colombi, S., Hivon, E., \& Juszkiewicz,
R. 1995, A\&A, 296, 575

\bibitem[Bouchet \etal\ 1992]{B92}
Bouchet, F. R., Juszkiewicz, R., Colombi, S., \& Pellat, R. 1992, \apj,
394, L5

\bibitem[Bouchet \etal\ 1993]{B93} 
Bouchet, F. R., Strauss, M. A.,Davis, M., Fisher, K. B., Yahil, A., \&
Huchra, J. P. 1993, \apj, 417, 36 (B93)

\bibitem[]{10} Coles, P., \& Jones, B. J. T. 1991, MNRAS, 248, 1

\bibitem[]{11}
Colombi, S., Bernardeau, F., Bouchet, F. R., \& Hernquist, L. 1997,
MNRAS, 287, 241

\bibitem[]{CBH}
Colombi, S., Bouchet, F. R., \& Hernquist, L. 1996, \apj, 465, 14 

\bibitem[]{12}
Colombi, S., Bouchet, F. R. \& Schaeffer, R. 1994, \aap, 281, 301

\bibitem[]{13}
Colombi, S., Bouchet, F. R. \& Schaeffer, R. 1995, ApJS, 96, 401

\bibitem[]{14} 
Cram\'er, H. 1946, Mathematical Methods of Statistics (Princeton:
Princeton Univ. Press)

\bibitem[]{15} da Costa, L. N., Pellegrini, P., Davis, M., Meiksin, A.,
Sargent, W., \& Tonry, J. 1991, ApJS, 75, 935

\bibitem[]{16}
Fisher, K., B., Davis, M., Strauss, M., A., Yahil, A., Huchra, J., P. 
1993, \apj, 402, 42

\bibitem[]{17} Fisher, K. B., Huchra, J. P., Davis, M., Strauss, M. A.,
Yahil, A., \& Schlegel, D. 1995, ApJS, 100, 69

\bibitem[]{18} Fry, J. N. 1984a, ApJ, 277, L5

\bibitem[Fry 1984]{F84}
Fry, J. N. 1984b, \apj, 279, 499

\bibitem[]{19}
Fry, J. N., \& Gazta\~naga, E. 1993, \apj, 413, 447

\bibitem[]{20}
Fry, J. N., \& Gazta\~naga, E. 1994, \apj, 425, 1

\bibitem[]{21} 
Fry, J. N., \& Scherrer, R. J. 1994, ApJ, 429, 36

\bibitem[]{22}
Gazta\~naga, E. 1992, \apj, 398, L17

\bibitem[]{23}
Gazta\~naga, E. 1994, MNRAS, 268, 913

\bibitem[]{24}
Gazta\~naga, E. 1995, ApJ, 454, 561

\bibitem[]{25}
Gazta\~naga, E., \& Frieman, J. A. 1994, \apj, 437, L13

\bibitem[]{26}
Gazta\~naga, E., \& Yokoyama, J. 1993, \apj, 403, 450

\bibitem[]{27} Gunn, J. E., \& Weinberg, D. H. 1995, in {\it Wide-Field
Spectroscopy and the Distant Universe}, ed.\ S. J. Maddox and A.
Arag\'on-Salamanca (Singapore: World Scientific), 3

\bibitem[]{28} Hivon, E., Bouchet, F. R., Colombi, S., \& Juszkiewicz,
R. 1995, A\&A, 298, 643

\bibitem[]{29} Huchra, J. P., Davis, M., Latham, D., \& Tonry, J. 1983,
ApJS, 52, 89

\bibitem[Juszkiewicz, Bouchet \& Colombi 1993]{JBC}
Juszkiewicz, R., Bouchet, F. R., \& Colombi, S. 1993, \apj, 412, L9

\bibitem[Juszkiewicz \etal\ 1995]{J95}
Juszkiewicz, R., Weinberg, D. H., Amsterdamski, P., Chodorowski, M.,
\& Bouchet, F. R. 1995, \apj, 442, 39

\bibitem[]{30}
Kofman, L., Bertschinger, E., Gelb, J. M., Nusser, A., \& Dekel,
A. 1994, \apj, 420, 44
 
\bibitem[]{31} Lahav, O., Itoh, M., Inagaki, S., \& Suto, Y. 1993, ApJ,
402, 387

\bibitem[{\L}okas \etal\ 1995]{l95}
{\L}okas, E. L., Juszkiewicz, R., Weinberg, D. H., \& Bouchet, F. R. 
1995, MNRAS, 274, 730

\bibitem[]{32} Meiksin, A., Szapudi, I., \& Szalay, A. S. 1992, ApJ, 394, 87

\bibitem[]{33} Park, C. 1990, PhD Thesis, Princeton University

\bibitem[Peebles 1980]{P80}
Peebles, P. J. E. 1980, {\it The Large-Scale Structure of the Universe}
(Princeton Univ. Press)

\bibitem[]{34} Press, W. H., Flannery, B. P., Teukolsky, S. A., \&
Vetterling, W. H. 1992, Numerical Recipes, The Art of Scientific
Computing (Second Edition) (Cambridge: Cambridge University Press)

\bibitem[]{35} Protogeros, Z. A. M., \& Weinberg, D. H. 1997, ApJ,
submitted (astro-ph/9701147)

\bibitem[]{36} Scherrer, R. J., \& Bertschinger, E. 1991, ApJ, 381, 349

\bibitem[]{37}
Strauss, M. A., Davis, M., Yahil, A., \& Huchra, J. P. 1992, \apj,
385, 421

\bibitem[Strauss & Willick]{SW}
Strauss, M. A., \& Willick, J. A. 1995, Physics Reports, 261, 271 

\bibitem[]{38}
Szapudi, I., \& Colombi, S. 1996, ApJ, 470, 131

\bibitem[]{39} Szapudi, I., Dalton, G. B., Efstathiou, G., \& Szalay, A. 
S. 1995, ApJ, 444, 520

\bibitem[]{40} Szapudi, I., Meiksin, A., \& Nichol, R. C. 1996, ApJ, 473, 15

\bibitem[]{41} Szapudi, I., \& Szalay, A. S. 1996, ApJ, 459, 504

\bibitem[]{42} Szapudi, I., Szalay, A. S., \& Boschan, P. 1992, ApJ, 390, 350

\bibitem[]{43} Ueda, H., \& Yokoyama, J. 1996, MNRAS, 280, 754 

\bibitem[]{44} Zel'dovich, Y. B. 1970, A\&A, 5, 84

\end{thebibliography}
\end{document}